%% file: main.tex
\documentclass[fleqn,usenatbib]{mnras}
\usepackage{bigints}
\usepackage{cancel}
\usepackage{footnote}
\usepackage{amsmath}
\usepackage[dvipsnames]{xcolor}
\makesavenoteenv{tabular}
\usepackage[normalem]{ulem}
\usepackage{tabulary}
\usepackage{colortbl}
\usepackage{amssymb,amsmath, amstext}
\usepackage{graphicx}
\usepackage[T1]{fontenc}
\usepackage{url}
\usepackage{newtxtext,newtxmath}
\usepackage{bm}
\usepackage{comment}
\usepackage{color}
\usepackage{listings}
\usepackage{natbib}
\usepackage[capitalise]{cleveref}
\usepackage{hyperref}
\usepackage{tikz}
\usepackage{nicematrix}
\usepackage{siunitx}

\usepackage{booktabs}
\begin{document}
\makeatletter
\tikzset{
    double color fill/.code 2 args={
        \pgfdeclareverticalshading[%
            tikz@axis@top,tikz@axis@middle,tikz@axis@bottom%
        ]{diagonalfill}{100bp}{%
            color(0bp)=(tikz@axis@bottom);
            color(50bp)=(tikz@axis@bottom);
            color(50bp)=(tikz@axis@middle);
            color(50bp)=(tikz@axis@top);
            color(100bp)=(tikz@axis@top)
        }
        \tikzset{shade, left color=#1, right color=#2, shading=diagonalfill}
    }
}
\makeatother

\tikzset{%
  diagonal fill/.style 2 args={%
      double color fill={#1}{#2},
      shading angle=45,
      opacity=0.8},
  other filling/.style={%
      shade,
      shading=myshade, 
      shading angle=0,
      opacity=0.5}
 }

\pgfdeclarehorizontalshading{myshade}{100bp}{%
    color(0bp)=(blue);
    color(25bp)=(blue);
    color(37.5bp)=(blue);
    color(37.5bp)=(brown);
    color(50bp)=(brown);
    color(50bp)=(green);
    color(62.5bp)=(green);
    color(62.5bp)=(purple);
    color(75bp)=(red);
    color(100bp)=(red)}

\newcommand{\corr}[1]{{\bf\textcolor{purple}{ #1}}}
\newcommand{\corrceline}[1]{{\bf\textcolor{orange}{ #1}}}
\newcommand{\corrmariana}[1]{{\bf\textcolor{green}{ #1}}}

\title[Unbinned likelihood with super-sample covariance]{Towards including super-sample covariance in the unbinned likelihood for cluster abundance cosmology}
\author[Payerne, Murray, Combet, Penna-Lima]{
\parbox{\textwidth}{
\Large
C. Payerne,$^{1,2}$\thanks{E-mail: constantin.payerne@cea.fr}
C. Murray,$^{2,3}$
C. Combet,$^{2}$
M. Penna-Lima$^{4, 5, 6}$
}
\vspace{0.4cm}
\\
\parbox{\textwidth}{
$^{1}$ Université Paris-Saclay, CEA, IRFU, 91191 Gif-sur-Yvette, France\\
$^{2}$ Université Grenoble Alpes, CNRS-IN2P3, LPSC, 38000 Grenoble, France\\
$^{3}$ Université Paris Cité, CNRS-IN2P3, APC, 75013 Paris, France\\
$^{4}$ Instituto de Física, Universidade de Brasília, 70910-900, Brasília, DF, Brazil \\
$^{5}$ Centro Internacional de F\'isica, Instituto de F\'isica, Universidade de Bras\'ilia, 70910-900, Bras\'ilia, DF, Brasil \\
$^{6}$ Departamento de F\'isica, Universidade Estadual de Londrina, 
Rod. Celso Garcia Cid, Km 380, 86057-970, Londrina, Paran\'a, Brazil
}
}
\pubyear{2024}
\date{Accepted XXX. Received YYY; in original form ZZZ}
\maketitle
\begin{abstract}
The measurement of the abundance of galaxy clusters in the Universe is a sensitive probe of cosmology, which depends on both the expansion history of the Universe and the growth of structure. Density fluctuations across the finite survey volume add noise to this measurement, this is often referred to as super-sample covariance (SSC). For an unbinned cluster analysis, such noise has not been included in the cluster likelihood, since the effect of SSC was small compared to the Poisson shot-noise for samples of a few hundred clusters. For upcoming large cluster surveys such as the Rubin LSST, which will deliver catalogs of tens of thousands of clusters, this effect will no longer be negligible. 
In this paper, we propose a new hybrid likelihood based on the Gauss-Poisson Compound model (GPC), by using infinitesimal mass bins and standard redshift bins. This likelihood has the advantages of an unbinned Poisson likelihood whilst successfully incorporating the effects of SSC. Using a simulated dark matter halo catalog, we find that the hybrid likelihood, accounting for both Poisson noise and SSC, increases the dispersion of the parameter posteriors by 20$\%$ when using 100,000 clusters compared to the standard unbinned likelihood, based on Poisson statistics only. 


\end{abstract}



\begin{table*}
    \centering
    \begin{tabular}{l||l}
     \hline
    Name/Acronyms & Full name/description \\
    \hline
         ULC& Unbinned Likelihood for Clusters \\
         HLC& Hybrid Likelihood for Clusters \\
         BLC& Binned Likelihood for Clusters \\
         \hline
         SSC & Super-Sample Covariance\\
         SN & Poisson shot noise\\
         Poisson & Poisson distribution\\
         Gauss-SN+SSC& Gaussian likelihood with covariance \\&accounting for SSC and SN \\
       GPC& Gauss-Poisson Compound\\
    \end{tabular}
    \caption{Table of acronyms and associated full names/definitions used in this work. The first block corresponds to the binning method for counting clusters, the second block corresponds to the statistical properties of counts. }
    \label{tab:table_likelihood}
\end{table*}
\section{Introduction}

Galaxy clusters are the largest gravitationally bound objects in the Universe and form through the gravitational collapse of the largest matter density fluctuations in the Universe. Their formation history, mass, and spatial distribution are highly sensitive to the fluctuations in the matter density field, the expansion rate of the Universe, and the nature of gravity (e.g., \citet{Bartlett1997}). Over the past two decades, the measurement of the abundance of galaxy clusters in the Universe has provided competitive constraints on the average matter density in the Universe $\Omega_m$ and the amplitude of matter density fluctuations $\sigma_8$ from optical cluster catalogs (e.g., \citet{Abbott2020DESCL,Lesci2022KIDSCL,Fumagalli2023SDSS}), X-ray catalogs (e.g., \citet{Mantz2014WTGCL,Ghirardini2024erositaCL}) and catalogs of clusters detected through the Sunyaev-Zeldovich (SZ) effect at millimeter wavelengths (e.g., \citet{Ade2016PlanckCL,Bocquet2024SPT}). 

Given a cosmological model and a measurement of the abundance of galaxy clusters, we can infer cosmological parameters from the observed cluster abundance by using two different likelihood methods: binned and unbinned. The binned approach consists of counting galaxy clusters in redshift and mass-proxy bins and can account for both the Poisson noise \citep{Poisson1837} and the super-sample covariance \citep{2003_SV_HU}, the latter is due to the coupling of small and large scale modes of the matter density field (both within and beyond the survey volume). This introduces a covariance between clusters within different mass and redshift bins and increases their variance within a bin.  For large-scale structure probes such as weak lensing, galaxy clustering, or galaxy clusters, the impact of SSC is significant and may increase the error on cosmological parameter constraints by as much as 20 $\%$ for large surveys \citep{Gouyou2022SSC,Fumagalli2021pinocchiovariance,Payerne2023}. For abundance-based analyses, the relative SSC contribution increases with the total number of clusters in the sample. It is therefore an important effect that has to be carefully taken into account when analyzing catalogs of a few thousand objects, such as provided by the Dark Energy Survey (DES, \citet{DES2005wpaper}) or the Kilo Degree Survey (KiDS, \citet{deJong2013kidswp}). The effect of SSC will be important for upcoming large cluster surveys such as the Legacy Survey of Space and Time of the Vera Rubin Observatory \citep{LSST}, the \textit{Euclid} mission \citep{laureijs2011euclid} and Simons Observatory \citep{S02019whitepaper} that will detect tens to hundreds of thousand clusters.

When working in a binned abundance framework and a sufficient number of clusters per bin, the SSC can be included in a Gaussian likelihood (see e.g. \citet{Abbott2020DESCL,Lesci2022KIDSCL}).
Alternatively, the unbinned framework permits to account for correlations between different cluster properties at the level of the cluster catalog.
Furthermore, we expect the unbinned likelihood to provide tighter constraints if the correlations between cluster properties are significant. For example in the binned approach of  \cite{Costanzi2019SDSSCL} and \cite{Fumagalli2023SDSS} their constraints are prior-dominated from their constraints on the intrinsic scatter between cluster richness and weak-lensing mass, whereas an unbinned approach can directly constrain this scatter (for example \citet{Bocquet2019SPTCL,Murray2022lensing} constrain the intrinsic scatter between cluster mass-proxies and weak lensing mass using individual weak lensing masses). 
However, the standard approach to the unbinned framework \citep{Mantz2010CCmethodunbinned} cannot account for the SSC at the level of the cluster catalog. In this work, we present a hybrid likelihood (binned in redshift but not in mass), that allows us to account for the SSC while preserving the benefits of the unbinned approach. This likelihood can be used when SSC becomes an important contribution to the overall abundance noise and one wishes to constrain cosmology with clusters at the catalog level.

We first review existing likelihoods for cluster abundance cosmology in \cref{sec:existing_cluster_count_lik}. Binned likelihoods are presented in \cref{sec:likelihoods_binned} and the standard unbinned likelihood based on Poisson statistics in \cref{sec:standard_unbinned_likelihood}. The various abundance-based cosmological analyses found in the literature are presented \cref{sec:litterature}. In \cref{sec:unbinned_SSC} we discuss how to account for SSC in an unbinned framework; the main result of this work is presented in \cref{eq:hybrid_likelihood_unbinned}, where we propose a new hybrid likelihood, by using infinitesimal mass bins and standard redshift bins. We use this hybrid likelihood in \cref{sec:pinochio} to test how the SSC contribution impacts the parameter posterior using PINOCCHIO dark matter halo catalogs before concluding in \cref{sec:conclusions}.

\begin{table*}
\begin{center}
\begin{tabular}{c||l|c|c|c|c}
\hline
Survey & Analysis & $\Omega_S$ (deg${}^2$) & $z_{\rm min}$ - $z_{\rm max}$ & $N_{\rm tot}$ & Likelihood 
  \\
 \hline
  ROSAT & WtG, \citet{Mantz2014WTGCL}&400&0 - 0.5&224 & ULC/Poisson\\
  XMM &XXV, \citet{Pacaud2018XXLCL} &50&0.05 - 1& 178&(ULC/Poisson)$\times$(BLC/GPC)\\
   &XLVI, \citet{Garrel2022XMM} &47.36&0.05 - 1& 178&BLC/GPC\\
eROSITA & eFEDS, \citet{Chiu2022erositaCL} &140&0.1 - 1.2& 455&ULC/Poisson\\
 & eRASS1, \citet{Ghirardini2024erositaCL} &12,791&0.1 - 0.8& 5,259&ULC/Poisson\\
 \hline
  ACT&S11, \citet{Sehgal2011actCL} &455&0.16 - 1&9 &BLC/Poisson\\
 &H13, \citet{Hasselfield2013ACTCL} &504&0.15 - 0.8&15 &ULC/Poisson\\
 \textit{Planck}&XX, \citet{Ade2014PlanckCL}&26,000&0 - 1&189&BLC/Poisson\\
&XXIV, \citet{Ade2016PlanckCL}&26,000&0 - 1&439&BLC/Poisson\\
&S18, \citet{Salvati2018Planck}&26,000&0 - 1&439&BLC/Poisson\\
&Z19, \cite{Zubeldia2019PlanckCMBlensing} & 26,000 & 0 - 1 &439 &ULC/Poisson \\
&$\times$SPT, \citet{Salvati2022PlanckSPT}& (26,000)$\times$(2,500)&0 - 1&782&(ULC/Poisson)$\times$(BLC/Poisson)\\
&$\times$\textit{Chandra}, \cite{Aymerich2024PlanckChandraCL} & 26,000 & 0 - 1 &439 &BLC/Poisson\\
&$\times$ACT, \cite{Lee2024PlanckACTCL} & (26,000)$\times$(987.5) & 0 - 1.4 &439 &(BLC/Poisson)$\times$(ULC/Poisson)\\
SPT&B15, \citet{Bocquet2015SPTCL}&720&$z > 0.3$&100&ULC/Poisson\\
 &H16, \citet{deHann2016SPTCL}&2,500&0.25 - 1.7&377&ULC/Poisson\\
  &B19, \citet{Bocquet2019SPTCL} &2,500&0.25 - 1.75&343 &ULC/Poisson\\
  &C22, \citet{Chaubal2022SPTCLCMBlensing} &2,500&0.25 - 1.75&343 &ULC/Poisson\\
  &B24, \citet{Bocquet2024SPT} &5,200&0.25 - 1.75&1,005 &ULC/Poisson\\
\hline
 SDSS&R10, \citet{Rozo2010CLSDSS}&7,398&0.1 - 0.3&10,810&BLC/Gauss-SN+SSC\\
   &M13, \citet{Mana2013SDSScosmologyCL}&7,500&0.1 - 0.3&13,823&BLC/Gauss-SN+SSC\\
&C19, \citet{Costanzi2019SDSSCL} &11,000&0.1 - 0.3&7,000&BLC/Gauss-SN+SSC\\
 &A20, \citet{Abdullah2020SDSSCL}&11,000&0.045 - 0.125&756&BLC/Gauss-SN+SSC\\
 &P23, \citet{Park2023lensingabundance}&11,000&0.1 - 0.33&8,379&BLC/Gauss-SN+SSC\\
 &S23, \citet{Sunayama2023HSClensing}&11,000&0.1 - 0.33&8,379&BLC/Gauss-SN+SSC\\
&F23, \citet{Fumagalli2023SDSS}&11,000&0.1 - 0.3&6,964&BLC/Gauss-SN+SSC\\
DES &Y1, \citet{Abbott2020DESCL} &1,800&0.2 - 0.65&7,000&BLC/Gauss-SN+SSC\\
 &T21, \citet{To2021DEScomb} &1,321&0.2 - 0.6&4,794&BLC/Gauss-SN+SSC\\
    &$\times$SPT, \citet{Costanzi2021DESSPTCL} & (1,800)$\times$(2,500)& $z > 0.2$ &7,000&BLC/Gauss-SN+SSC\\
  KiDS&DR3, \citet{Lesci2022KIDSCL} &377&0.1 - 0.6&3,652& BLC/GPC\\
\end{tabular}
\caption{An overview of abundance-based cosmological analyses. X-ray surveys are in the first horizontal block, millimeter surveys in the second block, and optical surveys in the third block. The first column gives the name of the survey on which the clusters have been detected, the second column indicates the different analyses. The third column indicates the sky coverage, $\Omega_S$, of the survey in square degrees, the fourth column is the redshift range spanned by the cluster catalog, and the fifth gives the total number of clusters used in the analysis. Finally, the last column indicates the type of likelihood that used in the analysis. These acronyms are detailed in \cref{tab:table_likelihood}.}
\label{table:state_of_the_art}
\end{center}
\end{table*}
\section{Review of existing cluster count likelihoods}
\label{sec:existing_cluster_count_lik}
\subsection{Binned likelihoods}
\label{sec:likelihoods_binned}

The binned approach consists of considering a fixed binning scheme of the redshift-proxy plane and counting tens, hundreds, or thousands of clusters in each bin. The measured counts $\widehat{N}_k$, where $k$ denotes a given 2D mass-redshift bin $(\Delta m, \Delta z)$, are then compared to the predicted counts in bins,
\begin{equation}
    N_k = \int_{\Delta z}\int_{\Delta m}\Omega_S\frac{dn(m, z)}{dm}\frac{d^2V(z)}{dzd\Omega} dz dm,
    \label{eq:abundance_binned}
\end{equation}
where $dn(m,z)/dm$ is the mass function of objects at redshift $z$ and mass $m$, $\Omega_S$ is the survey solid angle and $V(z)$ the comoving volume. A counting experiment is affected by several sources of noise. The first is the Poisson noise, characteristic of a counting experiment of discrete objects in bins. The counts are considered to be uncorrelated, and each count has a variance called the Poisson shot noise equal to $N_k$. The second source of noise is the super-sample covariance, arising from fluctuations of the underlying matter density field, from which clusters are formed. Due to non-linear couplings between the different modes of the matter density fields (within and beyond the survey volume), the clustering of matter can be decreased or increased within the survey volume, which in turn leads to a decrease or increase in the observed number of clusters. This effect introduces correlations between the cluster counts in different bins of mass, redshift, and spatial position. It typically depends on the survey geometry, as well as the product of the counts of clusters $N_k$ and $N_l$ in bins $k$ and $l$ denoted as $N_k N_l$. Although ignored within this work, a third contribution to the scatter of counts is halo exclusion, this accounts for the effect that two halos cannot exist in the same spatial location. This will also increase the scatter of cluster counts (e.g. see \citet{Philcox2020haloexclusion}). Other sources of scatter and bias in count statistics arise from the uncertainties in the halo mass determination, the intrinsic scatter of cluster proxy, cluster photometric redshift errors, etc. (see e.g. \citet{Abbott2020DESCL}),  which is not the focus of this work. 

Several binned likelihoods have been used in the literature. When the cluster sample is small (of a few hundred objects), the Poisson shot noise is the dominant source of scatter and counts are described by a Poisson likelihood. Conversely, when cluster samples are large (comprising thousands, tens of thousands, or even hundreds of thousand objects) the effect of SSC becomes comparable to the Poisson shot noise. In this case, the cluster counts can be described by a Gaussian likelihood, with a covariance matrix given by the sum of the Poisson variance and the SSC. A third likelihood, the Gauss-Poisson Compound likelihood (GPC, see e.g. \citet{2003_SV_HU,Lima2024GPC,Hu_2006,Smith2011GPC,Payerne2023}), provides a more accurate description of cluster count than either the Poisson or the Gaussian distribution. GPC describes cluster counts as a result of a Poisson process, for which the Poisson mean is given by the cosmological expectation $N_k$ and a fluctuation $\delta N_k$, arising from the stochastic properties of the underlying matter density field, described by the SSC effect. For a cluster count experiment in a given mass and redshift bin, the associated total likelihood is obtained by averaging over all Poisson likelihoods with respective mean $N_k + \delta N_k$. As a result, the Poisson likelihood is the low count limit of the GPC likelihood, whereas the Gaussian is the large count limit, where SSC effects are no longer negligible. These three binned likelihoods are presented in detail in \cite{Hu_2006,Payerne2023}. 
\subsection{Standard unbinned likelihood}
\label{sec:standard_unbinned_likelihood}
The unbinned approach uses the information at the individual cluster level, such as individual cluster redshifts and mass proxies \citep{Cash1979unbinned,Mantz2010CCmethodunbinned, PennaLima_thesis,2014_PennaLima}. Therefore, it is sensitive to the full shape of the cluster mass-redshift distribution, contrary to the binned approach. However, as the halo mass function has a simple shape, most of this information can be recovered with relatively few bins (as shown in \citet{Payerne2023}).

Where the unbinned likelihood proves particularly useful is for studying the correlations between different observable properties of individual galaxy clusters; such as the cluster richness, weak lensing mass, X-ray, and SZ masses (see e.g. \citet{Mahdavi2013XMM,Murray2022lensing}).  \cite{Chiu2022erositaCL} and \cite{Bocquet2023SPTDES} have shown that the unbinned framework can be used to constrain the uncertainty and correlation between these different cluster properties (see \citet{Evrard2014multi} for a detailed description of the effects of correlation between different cluster properties). Moreover, the unbinned framework is useful to extract cosmological information from relatively small cluster samples (of a few hundred). It is however more computationally demanding than a binned approach since cosmological quantities need to be evaluated for each cluster. The computational cost is even larger when accounting for observational uncertainties.

The unbinned count regime is obtained by considering each bin of the mass-redshift binning scheme to be small enough to contain at most one cluster per bin. 
Let us first consider a binning scheme in the mass-redshift plane. The unbinned regime considers that the mass-redshift bins become infinitesimally small, such that the $k-$th mass-redshift bin centered on $m_k$ and $z_k$ has a \textit{surface} in the mass-redshift space of $\Delta z\Delta m$. Let $N_k$ be the predicted abundance within the small area $\Delta z \Delta m$ that is given from \cref{eq:abundance_binned} by
\begin{equation}
\label{eq:muk_approx_low_abundance}
    N_k \approx \Omega_S\frac{dn(m_k, z_k)}{dm}\frac{d^2V(z_k)}{dzd\Omega} \Delta z \Delta m = n_h(m_k,z_k) \Delta z \Delta m,
\end{equation}
where $\Delta m, \Delta z \ll 1$. Here, $n_h(m,z)$ is the predicted total halo number density per mass and redshift range. Considering Poisson statistics, the probability of the observed count $\widehat{N}_k$ is given by the Poisson law $\mathcal{P}(\cdot|N_k)$, 
\begin{equation}
  \mathcal{P}(\widehat{N}_k|N_k) =
    \begin{cases}
    e^{-N_k}=p_0 & \text{if $\widehat{N}_k = 0$},\\
      N_ke^{-N_k}=p_1 & \text{if $\widehat{N}_k = 1$},
    \end{cases}       
    \label{eq:P_Nk_unbinned}
\end{equation}
such that $p_0 + p_1=e^{-N_k}(1 + N_k) \approx 1$ if $N_k \ll 1$. Then, the observed count $\widehat{N}_k$ in that bin can only take two possible values $\{0, 1\}$, with respective probability $\{p_0, p_1\}$. Then, given a survey of $\widehat{N}_{\rm tot}$ clusters with masses $m_k$ and redshifts $z_k$, the Poisson unbinned likelihood is given as a product of individual likelihoods for each mass-redshift bin with index $k$, whether they are filled or not by at most one halo. We can derive the standard unbinned likelihood through  (see e.g. \citet{Cash1979unbinned})
\begin{equation}
    \mathcal{L}_{\rm ULC}^{\rm Poisson} = \prod_{k = 1}^{\rm all\ bins}\mathcal{P}(\widehat{N}_k|N_k) = 
    e^{-N_{\rm th}}\prod_{k = 1}^{\widehat{N}_{\rm tot}} N_k,
    \label{eq:P_unbinned_Poisson}
\end{equation}
by using \cref{eq:P_Nk_unbinned}, and where $N_{\rm th}$ is the predicted total cluster abundance at given cosmology, mass, and redshift ranges. 

\subsection{Literature overview}
\label{sec:litterature}

In this section, we present a literature overview of abundance-based cosmological analyses over the past 15 years. \cref{table:state_of_the_art} shows an overview of cluster abundance analyses using e.g. X-ray detected clusters by ROSAT \citep{Truemper1993rosatwp}, XMM \citep{xxl2016wp} or eROSITA \citep{Merloni2012erositaWP}, SZ clusters as detected by the \textit{Planck} satellite \citep{Tauber2010planckwp}, the Atacama Cosmology Telescope (ACT, \citet{Fowler2007actwp}), and the South Pole Telescope (SPT, \citet{Carlstrom2011SPTwp}) as well as optical clusters detected by the Sloan Digital Sky Survey (SDSS, \citet{York2000sdsswp}), the Dark Energy Survey (DES, \citet{DES2005wpaper}) and the Kilo Degree Survey (KiDS, \citet{deJong2013kidswp}).

These cluster abundance-based studies are usually combined with other cluster probes to calibrate the mass-proxy relation, one of the main limiting factors of abundance-based analyses. For optical/IR cluster surveys, the calibration is usually performed using weak-gravitational lensing of background galaxies (see e.g., DES, SDSS, or KiDS analyses). Weak lensing mass calibration has also been used for X-ray/SZ cluster surveys thanks to their overlap with optical/IR galaxy surveys, for instance, the recent \citet{Bocquet2024SPT} analysis combined the abundance of SPT-SZ clusters with the lensing signal estimated from DES and Hubble Space Telescope (HST) data, and \citet{Ghirardini2024erositaCL} combined eROSITA clusters with DES, Hyper Suprime-Cam (HSC) and KiDS lensing. Under the hydro-static equilibrium, cluster masses can also be constrained using the X-ray luminosity of the hot electron gas in galaxy clusters (see e.g. \citet{Ettori2013Xmass}) or combining X-ray and SZ cluster observations, the latter revealing the pressure of the hot electron gas \citep{Echeverria2023SZmass}.

Other works have made use of two different cluster abundance datasets, e.g. \citet{Salvati2022PlanckSPT} that constrained cosmology with the abundance of 439 \textit{Planck} clusters within $0 < z < 1$ (full-sky observations) and also with 343 SPT clusters between $0.25 < z < 1$ over 2,500 deg${}^2$ of the southern sky, and \citet{Costanzi2021DESSPTCL} that used $\sim 7000$ redMaPPer clusters from DES between $0.2 < z < 0.65$, and 141 SPT clusters with $z > 0.65$.

The last column of \cref{table:state_of_the_art} shows the different likelihoods that were used in the cosmological analyses, binned likelihoods are labeled as BLC for \textit{Binned Likelihood for Clusters} and unbinned likelihoods as ULC for \textit{Unbinned Likelihood for Clusters} (\cref{tab:table_likelihood} provides an overview and the definitions of the different acronyms used in the likelihood column in \cref{table:state_of_the_art} and in this work). From this table, we note that binned analyses based on a few hundred clusters (such as X-rays and SZ cluster samples) use a Poisson likelihood since the SSC is negligible compared to Poisson noise. Whereas, analyses of a few thousand clusters (e.g. SDSS, DES, and KiDS) account for both Poisson shot noise and SSC using a Gaussian likelihood. The more accurate GPC likelihood, valid in both Poisson and SSC regimes, was used to analyze the abundance of 178 XMM clusters and 3,652 KiDS clusters. 

\section{Including super-sample covariance in the unbinned cluster count likelihood}
\label{sec:unbinned_SSC}
\begin{figure*}
\begin{center}
\includegraphics[width=0.99\textwidth]{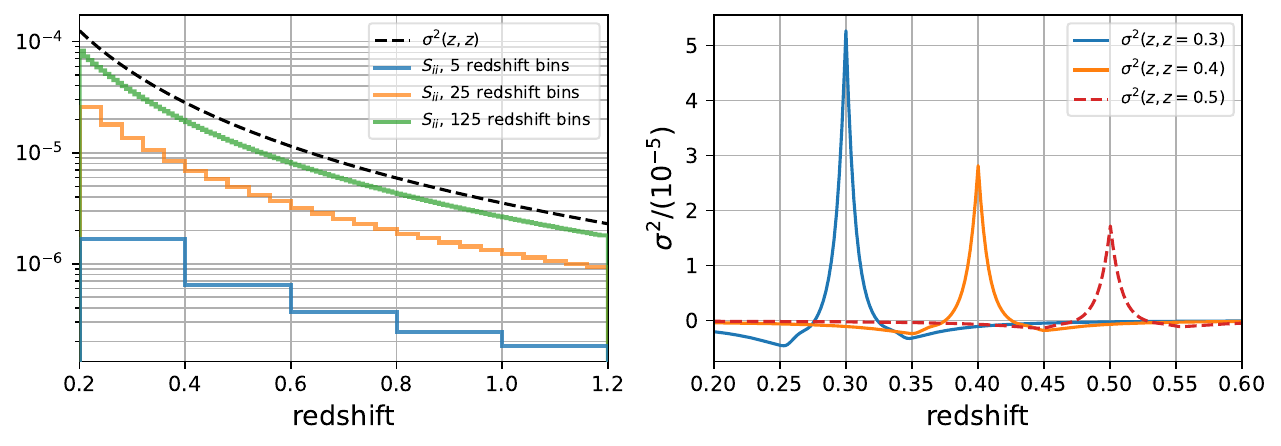}
\caption{Left: $\sigma^2(z,z)$ as a function of $z$ and binned variance $S_{ii}$ for different number of redshift bins in the redshift range $[0.2, 1.2]$. Right: $\sigma^2(z, z_1)$ as a function of $z$, for three different values of $z_1$ in the redshift range $[0.2, 0.6]$. All functions are plotted for the cosmology in \citet{Planck2014cosmology}. }
\label{fig:sigma_Sigma}
\end{center}
\end{figure*} 

The cluster likelihood in \cref{eq:P_unbinned_Poisson} is based on Poisson statistics, and does not account for the additive variance and correlation between counts due to the SSC. The latter is accounted for in binned analyses by considering an extra covariance term in the Gaussian likelihood or through a binned GPC likelihood. It is not straightforward to consider this effect at the unbinned level; it is the objective of this section.

Based on \citet{Mantz2010CCmethodunbinned} who proposed a joint likelihood $\mathcal{L}(\widehat{N}_{\rm tot}, \mathcal{O})$ for the total number of clusters $\widehat{N}_{\rm tot}$ and the set of individual observations $\mathcal{O}=\{m_k, z_k\}$ for $1\leq k \leq \widehat{N}_{\rm tot}$ (see details in \cref{app:pacaud}), \citet{Pacaud2018XXLCL} have proposed to change the Poisson distribution for the total count $\widehat{N}_{\rm tot}$ for a single-variate GPC likelihood, such that $\widehat{N}_{\rm tot}$ is sensitive to both Poisson noise and SSC. Therefore they have proposed to account for SSC in the unbinned likelihood based on a methodology that is relatively different from the unbinned count description proposed by \citet{Cash1979unbinned}, the latter we consider in this work (we discuss the main differences between the \citet{Mantz2010CCmethodunbinned} and \citet{Cash1979unbinned} prescriptions in \cref{app:pacaud}).

The focus of this section is to use the GPC description and the unbinned prescription proposed by \citet{Cash1979unbinned} to derive the likelihood $\mathcal{L}(\{\widehat{N}_{k}\})$ for $1\leq k \leq n_b$, instead of $\mathcal{L}(\widehat{N}_{\rm tot}, \mathcal{O})$. 
We first consider the general indexing $k$ for the $k$-th mass-redshift bin in a list of $n_b$ bins. For a set of observed cluster counts $\{ \widehat{N}_i \}$ and a fixed overdensity realization $\{\delta_{k}\}$ in each mass-redshift bin, the abundance likelihood is given by,

\begin{equation}
\label{eq:GPC_likelihood_multi}
    \mathcal{L}(\{\widehat{N}_{i}\}|\{x_k\}) = \prod_{k = 1}^{n_b}\mathcal{P}(\widehat{N}_k|x_k),
\end{equation}

\noindent where $1 \leq i \leq n_b$, $1 \leq k \leq n_b$, and each $x_k(\delta_k)$ is the \textit{local} mean. Denoting $\delta_k$ the background change of the matter density field due to the SSC, the linear prescription for $x_k$ is given by $x_k = N_k(1 + b_k\delta_k)$ (see e.g. \citet{Cole1989clusering,Mo1996clustering,Lima2024GPC}), where $b_k$ is the halo bias and

\begin{equation}
    \delta_k = \frac{1}{V_k}\int d\Vec{x}\ W_k(\Vec{x})\ \delta(\Vec{x}),
\end{equation}

\noindent where $W_k$ is the spatial window function for the $k$-th redshift bin, $V_k = \int d\Vec{x} \ W_k(\Vec{x})$ and $\delta(\Vec{x})$ is the matter density contrast. Each Poisson distribution is given by 
\begin{equation}
    \mathcal{P}(\widehat{N}_k|x_k) = \frac{x_k^{\widehat{N}_k}}{\widehat{N}_k!}e^{-x_k}.
    \label{eq:Poisson_law}
\end{equation}

In the above equations, we consider that the over-densities $\{\delta_k\}$ follow the Gaussian probability density function $p_{\rm SSC}(\{\delta_k\}) = \mathcal{N}(\{\delta_k\}|0, S)$, where $S = \{S_{kl}\}$, and where $S_{kl} = \langle \delta_k\delta_l \rangle$ is the covariance between the over-densities within the bins $k$ and $l$. The matrix $S_{kl}$ only depends of the redshift ranges spanned by the $k$ and $l$ bins, and is given by (see e.g. \citet{Lacasa_2018})
\begin{equation}
     S_{kl} = \int \frac{dV_{k}dV_{l}}{V_{k}V_{l}}\sigma^2(z_k, z_l),
     \label{eq:S_ij}
\end{equation}
where $\sigma^2(z_k, z_l)$ is the (full sky) amplitude of matter density fluctuations given by
\begin{equation}
    \sigma^2(z_1, z_2) = \int \frac{k^2 dk}{2\pi^2} j_0(kw(z_1))j_0(kw(z_2))P_{mm}(k|z_1, z_2).
    \label{eq:sigma2_z1z2_fullsky}
\end{equation}
where $j_0(x) = \sin(x)/x$ is the zero-th order spherical Bessel function, and $P_{mm}(k|z_1, z_2)$ is the matter power spectrum. We compute\footnote{We use \texttt{PySSC} \citep{Lacasa19,Gouyou2022SSC} to compute $\sigma^2$ (we use the \texttt{PySSC} function \texttt{sigma2}$\_$\texttt{fullsky}) and the $S_{ij}$ terms (\texttt{Sij}$\_$\texttt{alt}$\_$\texttt{fullsky}). 
The code is available at \url{https://github.com/fabienlacasa/PySSC}.} in \cref{fig:sigma_Sigma} (left panel) the function $\sigma^2(z, z)$ as a function of $z$ (dashed line) and the diagonal components of the $S_{ij}$ matrix for three different binning schemes, from five to 125 bins within the redshift range $0.2 < z < 1.2$. The variance $\sigma^2(z, z)$ decreases with redshift and the average variance of the matter density field in each redshift bin $S_{ii}$ converges to $\sigma^2(z, z)$ as the number of redshift bins increases. The function $\sigma^2(z, z_1)$ is represented as a function of $z$ in \cref{fig:sigma_Sigma} (right panel) for three different values of $z_1 = \{0.3, 0.4, 0.5\}$. The function $\sigma^2(z, z_1)$ is positive around $z_1$ and peaks for $z = z_1$.
\begin{figure*}
\begin{center}
\includegraphics[width=.49\textwidth]{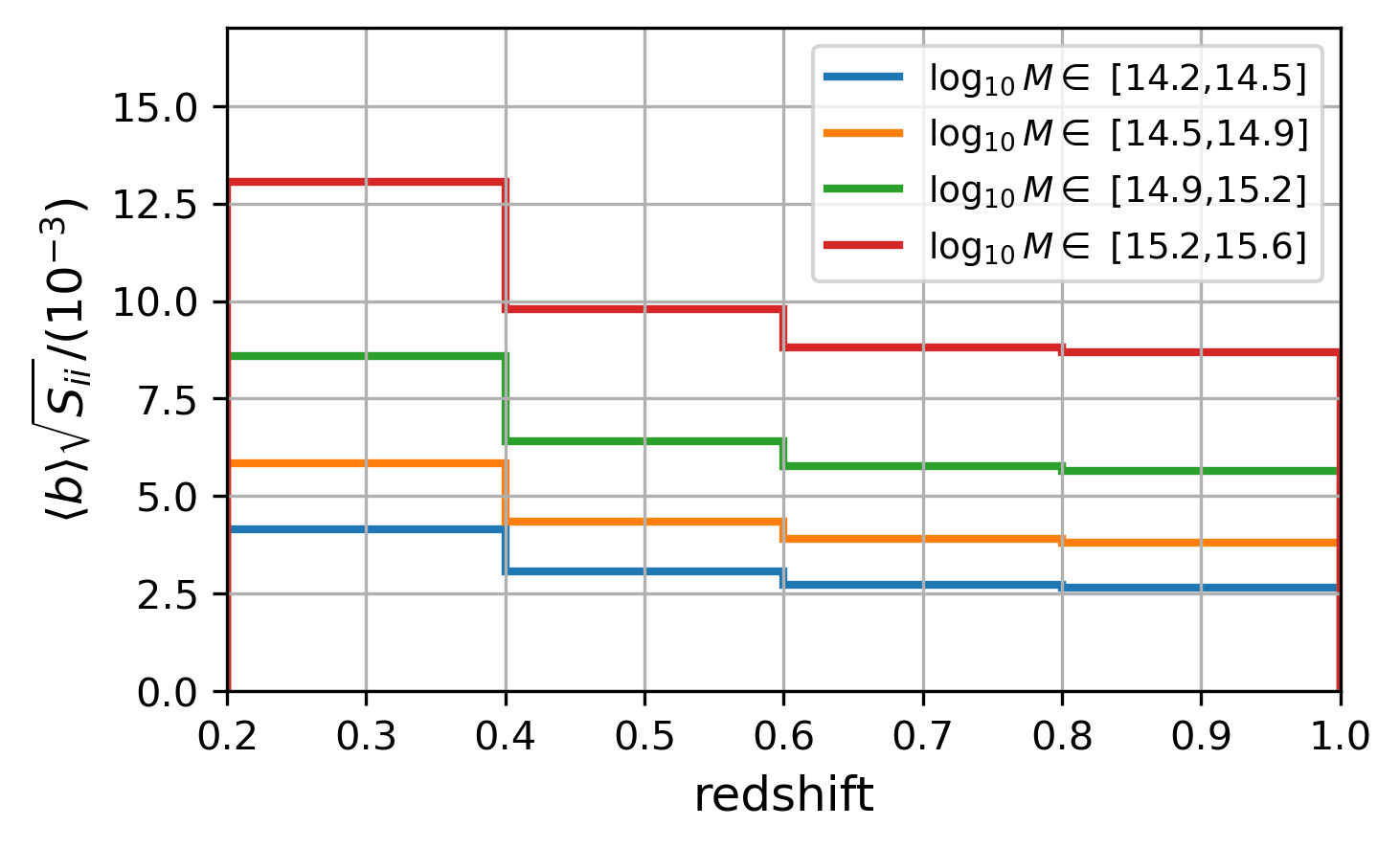}
\includegraphics[width=.49\textwidth]{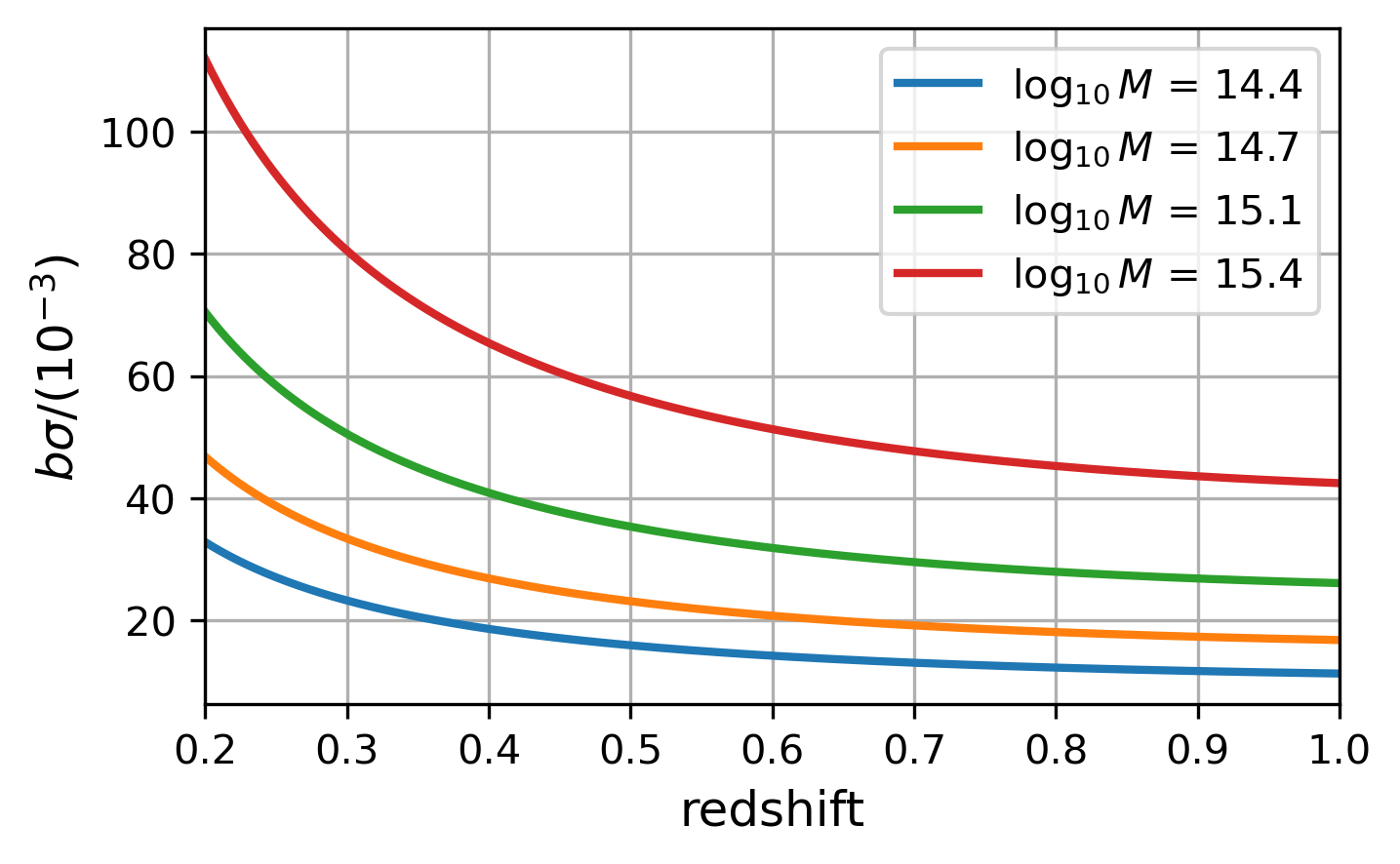}
\caption{Left: Average halo bias times the $S_{ii}^{1/2}$ elements (considering full-sky), as a function of different redshift bins and for different mass bins. Right:$b(m,z)\sigma_{\rm fullsky}(z,z)$ as a function of redshift, for different masses (halo bias from \citet{Tinker_2010}).}
\label{fig:diagnostic}
\end{center}
\end{figure*}

In terms of cosmological inference, the likelihood in \cref{eq:GPC_likelihood_multi} has $n_z$ new degrees of freedom\footnote{These are given by the over-densities $\{\delta_k\}_{1\leq k \leq n_z}$ in the $n_z$ redshift bins.} (in addition to the cosmological parameters such as $\Omega_m$ and $\sigma_8$) that can be considered as "nuisance parameters". \citet{Lacasa2023SSC} showed that, for large-scale structure probes, inferring cosmological parameters using the likelihood in \cref{eq:GPC_likelihood_multi} with $p_{\rm SSC}$ as a prior for the $\delta_k$ quantities (see e.g. \citet{Lesci2022KIDSCL}), is equivalent to using the average of the likelihood in \cref{eq:GPC_likelihood_multi} over the $p_{\rm SSC}$ distribution (see e.g. \citet{Garrel2022XMM,Payerne2023}). The second approach removes these extra degrees of freedom and this is the method we employ given that in the unbinned approach the number of bins becomes arbitrarily large.

The last step to obtain the unbinned likelihood is to consider infinitesimal mass and redshift bins, which are small enough to contain at most one cluster. The predicted abundance $N_{ik}$ in the narrow mass-redshift bin $ik$ given in \cref{eq:muk_approx_low_abundance}. This abundance regime needs to be applied to the \cref{eq:GPC_likelihood_multi}. 

We first have explored how to follow the binned-based approach developed by \citet{2014_Takada} in order to derive the unbinned likelihood with SSC. In their paper, they have proposed to make the Taylor expansion of the binned GPC model at the second order of matter density fluctuations $\delta_k$ under the two conditions $b_kN_k\delta_k \ll 1$ and $b_k\delta_k \ll 1$. Whereas these two conditions are met for the binned approach they have considered, the last one is not necessarily true when moving to smaller and smaller redshift bins. We note that, using the Taylor expansion approach to the second order, some caveats arises when using infinitesimal redshift bins. 
First, since the dispersion of $b_kN_k\delta_k$ given by $b(m_k,z_k)\sigma(z_k,z_k)$ becomes maximal in the unbinned regime, the perturbed local density $N_k(1+b_kN_k)$ can reach negative values, which leads to non-physical results (this effect is discussed in Appendix E of \citet{Lacasa19}, in the case of binned abundance analyses). This can be solved by using log-normal distributions  (see e.g. \citet{Coles1991density,Wen2020unbinnedPNL}), but this is difficult to implement. 
Second, limiting the Taylor expansion to the second-order terms may not be sufficient. Going to higher-order terms is a challenging derivation, which is not the focus of this work. We have summed up in \cref{app:W2} the calculations we have conducted in that direction, pointing again that they lead to nonphysical results.

We display in \cref{fig:diagnostic} (left panel) the variance of the \textit{binned} $b_k\delta_k$ quantity (considering with $f_{\rm sky} = 1$), to be compared to the unbinned version in the right panel. For the binned case, we consider 4 redshift bins from 0.2 to 1, and 4 mass bins in the $\log_{10}M$ range $[14.2, 15.6]$. The center of each mass bin is chosen such that the binned variance can be compared to the unbinned variance calculated in the right panel. In the left panel, the amplitude of $\langle b \rangle \sqrt{S_{ii}}$ is approximately 10 times lower in amplitude than that obtained for the unbinned regime (right panel). In other words, when considering larger redshift bins such as for the binned approach, the \textit{effective} variance of the halo overdensity $b_k\delta_k$ is significantly decreased e.g. the amplitude of matter fluctuations goes from $10^{-5}$ (unbinned) to $3\times10^{-7}$ (5 redshift bins) from right to left panel in \cref{fig:diagnostic}.

Then, using standard redshift bins in the likelihood rather than infinitesimal ones significantly decreases the variance of the volume-averaged matter density fluctuations, and thus allows to work around the difficulties mentioned above. Then, we introduce a new \textit{hybrid} likelihood (HLC for \textit{Hybrid Likelihood for Clusters}, see \cref{tab:table_likelihood}), with infinitesimal mass bins and a few fixed redshift bins (as for a binned approach).  Once the redshift binning is fixed, we can then keep the unbinned description for cluster masses and it is possible to build a \textit{hybrid} likelihood. The method to perform the new Taylor expansion following $b_k\delta_k \ll 1$ is based on the work led by \cite{Garrel2022XMM} (see their Appendix B). In our method, the count prediction $N_k$ is now different from what we have introduced in \cref{eq:P_unbinned_Poisson}; now, for a given $k$-th cluster with mass $m$ and within the redshift range $[z_m, z_{m+1}]$,
\begin{equation}
    N_k = dm_k\int_{\Delta z_m} dz\ n_h(z,m_k)  \ll 1.
    \label{eq:N_count_hybrid}
\end{equation}
Introducing $\langle \cdot\rangle_{\rm SSC}$ to be the average over all possible values of the redshift-binned overdensity $\delta_k$ (following the $p_{\rm SSC}$ distribution), we get the hybrid likelihood given by (see details in \cref{app:alternative_hybrid})
\begin{equation}
    \mathcal{L}_{\rm HLC}^{\rm GPC} = \mathcal{L}_{\rm HLC}^{\rm Poisson}\langle \mathcal{W}\rangle_{\rm SSC}.
\label{eq:hybrid_likelihood_unbinned}
\end{equation}
The $\mathcal{L}_{\rm HLC}^{\rm Poisson}$ term is the hybrid Poisson likelihood, by using $N_k$ in \cref{eq:N_count_hybrid} and the likelihood form in \cref{eq:P_unbinned_Poisson}. The $\langle\mathcal{W}\rangle_{\rm SSC}$ term denotes the contribution from SSC in the hybrid likelihood. Under the condition $\frac{1}{3}b_k^3\delta_k^3 \ll 1$, we obtain
\begin{equation}
    \langle \mathcal{W}\rangle_{\rm SSC} = \sqrt{\frac{\det S^{-1}}{\det C}}\exp\{\frac{1}{4}v^TC^{-1}v\},
    \label{eq:hybrid_Garrell}
\end{equation}
\begin{figure*}
\centering\includegraphics[width=.49\textwidth]{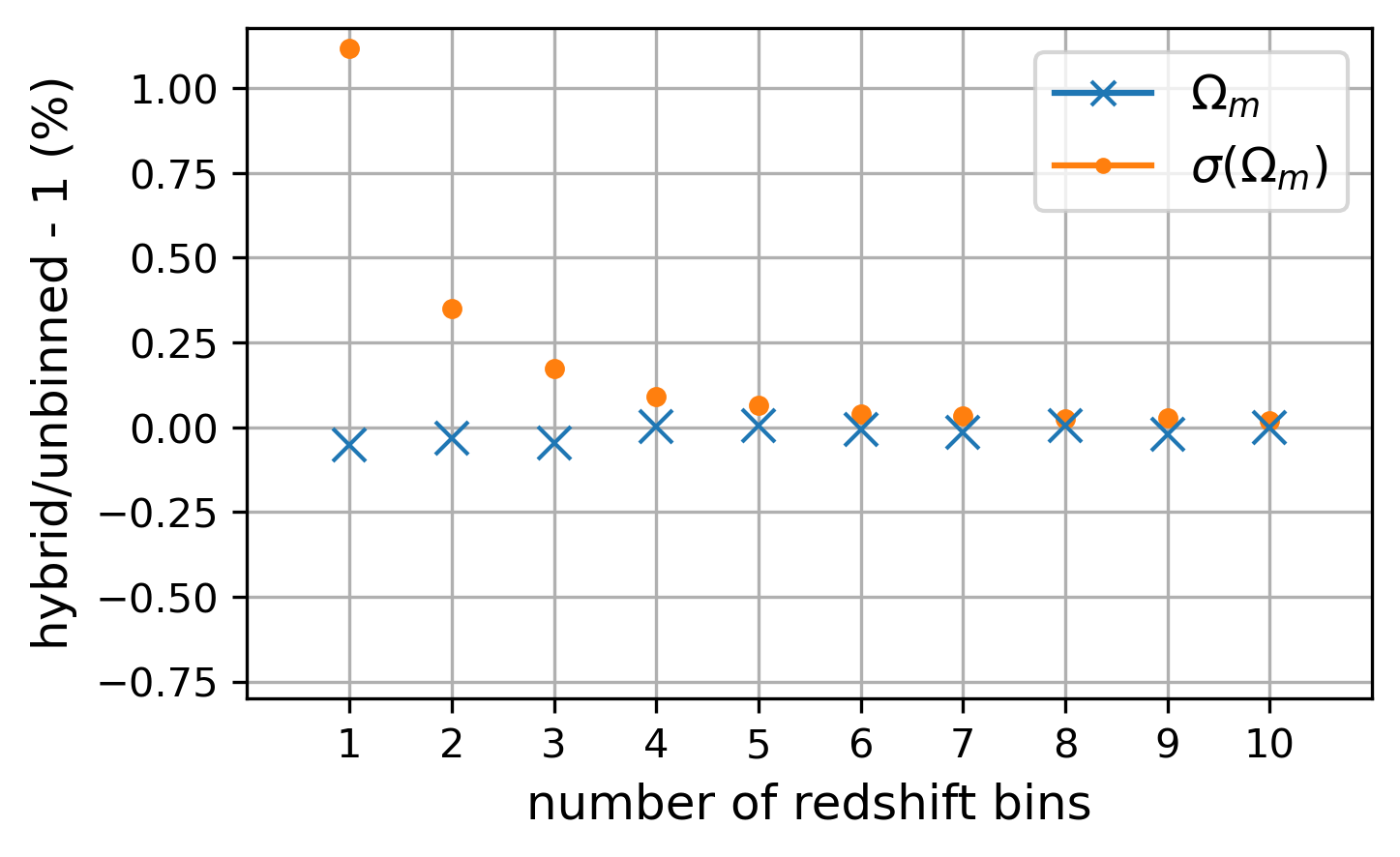}
\centering\includegraphics[width=.49\textwidth]{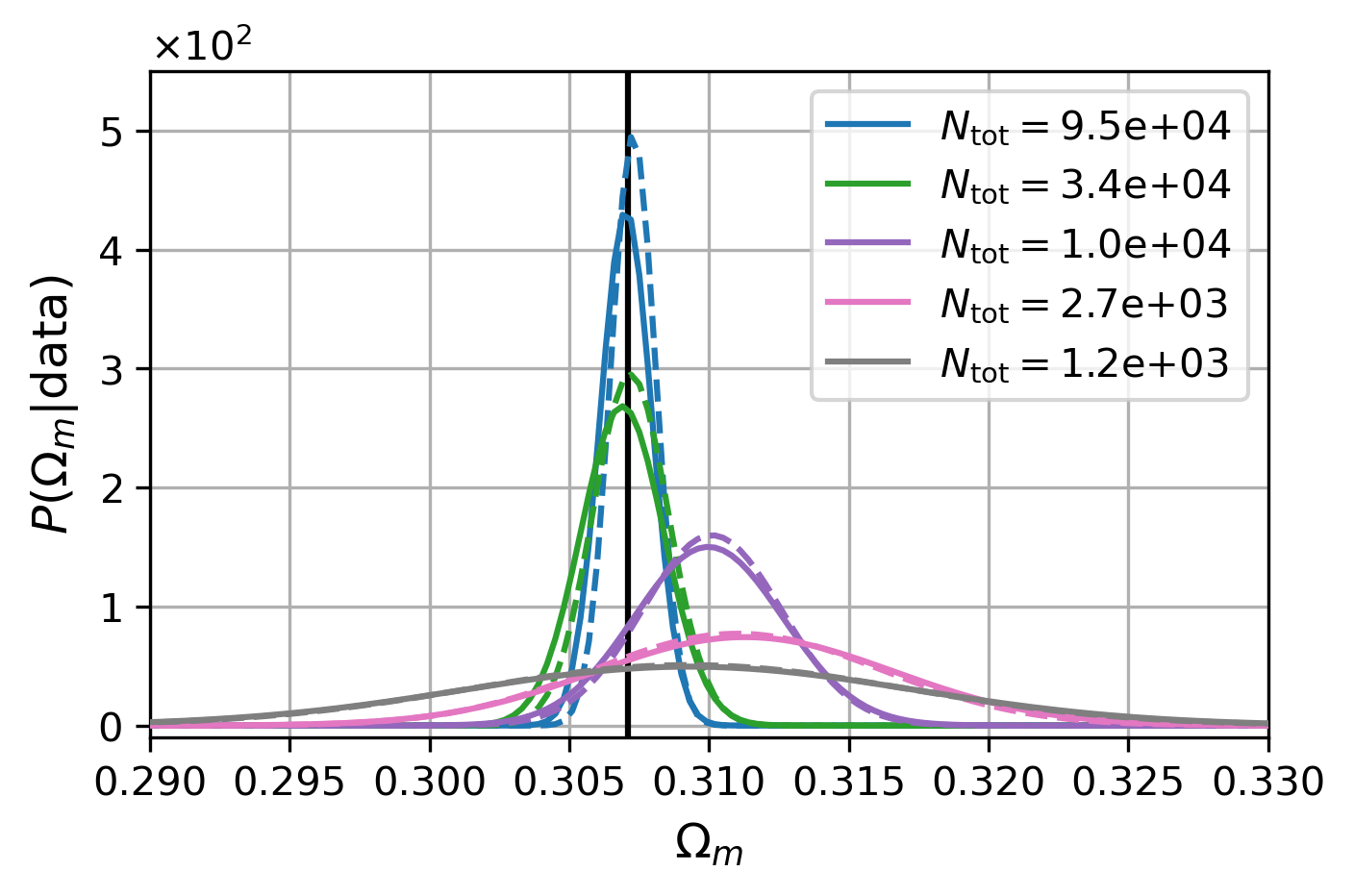}
    \caption{Left: Bias (in $\%$) on $\Omega_m$ and $\sigma(\Omega_m)$ when using the Poisson-only hybrid likelihood (i.e. by considering $\langle \mathcal{W}\rangle_{\rm SSC} = 1$ in \cref{eq:hybrid_likelihood_unbinned}) compared to the standard unbinned likelihood in \cref{eq:P_unbinned_Poisson}. We use the cluster sample with $\log_{10}M\in [14.2, 15.6]$ and with $z\in [0.2, 1.2]$, and the full redshift range is split in $n=1, 2, ..., 10$ equal-sized bins.  Right: Posterior distribution of $\Omega_m$ for different cluster samples for the mass range varying case (using $0.2 < z < 1.2$ and respectively using the mass cut $\log_{10}M > \log_{10}M_{\rm min} = \{14.2, 14.4, 14.6, 14.8, 14.9\}$. Dashed lines correspond to the Poisson-only hybrid likelihood (i.e. without SSC), and full lines correspond to the hybrid likelihood in \cref{eq:hybrid_likelihood_unbinned}. The vertical bar represents the input PINOCCHIO cosmology.}
    \label{fig:res_Omegam_posterior}
\end{figure*}
where $v_m = \langle Nb \rangle_m - \langle \widehat{Nb} \rangle_m$,
\begin{equation}
    [C]_{mn} = \frac{1}{2}[S_{mn}^{-1} + \langle\widehat{Nb^2} \rangle_n \delta_{mn}^K], 
\end{equation} and $\widehat{N}_{\rm tot}^m$ is the total number of clusters in the redshift bin $m$. Moreover, 
\begin{align}
\label{eq:Nb_m_th}
    \langle Nb \rangle_{m} &= \int_{\Delta m}\int_{\Delta z_m}dz\ dm\ n_h(m,z)b(m,z), \\
\label{eq:Nb_m_obs}
    \langle \widehat{Nb} \rangle_{m} &= \sum_{k=1}^{\widehat{N}_{\rm tot}^m}\widehat{N}_kb_m(m_k),\\
    \label{eq:N2b2_m_obs}
      \langle \widehat{Nb^2} \rangle_{m} &= \sum_{k=1}^{\widehat{N}_{\rm tot}^m}\widehat{N}_kb_m^2(m_k).
\end{align}
The observed cluster counts in each bin is $\widehat{N}_k = \{0,1\}$ such that $\widehat{N}_k=\widehat{N}_k^2$. The average bias in the redshift bin $[z_m, z_{m+1}]$ at mass $m_k$ is given by
\begin{equation}
    b_m(m_k) = \frac{1}{
    \int_{\Delta z_m}dz\ n_h(m_k, z)}\int_{\Delta z_m}dz\ n_h(m_k, z)  b(m_k,z). 
\end{equation}
As a result, \cref{eq:hybrid_likelihood_unbinned} is obtained without using the low abundance regime approximation $b_kN_k \ll 1$ that is necessary for the \citet{2014_Takada} approximation. In \cref{app:alternative_hybrid}, we present in more detail the condition for the validity of the hybrid likelihood, since it relies on the condition $b_k \delta_k \ll 1$; cluster catalogs obtained from surveys spanning a large sky fraction and large redshift range such as Rubin LSST, \textit{Euclid} or the Simons Observatory, will typically meet that condition.

For a single redshift bin denoted by the index $n$, we get a quite similar expression of what was found in \citet{Garrel2022XMM} (see their Eq. (7)) given by
\begin{align}
\begin{split}
    \langle \mathcal{W} \rangle_{\rm SSC} &= \left(1 + S_{nn}\langle \widehat{Nb^2}\rangle_n\right)^{-1/2}\times\\  \exp& \left( \frac{S_{nn}}{2}(\langle Nb \rangle_n - \langle \widehat{Nb} \rangle_n)^2\times(1 + S_{nn}\langle \widehat{Nb^2}\rangle_n)^{-1}\right).
\end{split}
\end{align}
In that sense, we have found an extension of the \citet{Garrel2022XMM} binned likelihood by (i) adding all possible correlations between redshift bins and (ii) defining the hybrid regime. 

\section{Impact of super-sample covariance on the hybrid cosmological inference}
\label{sec:pinochio}
In this section, we aim to estimate the impact of SSC on the parameter posterior given a cosmological dataset. We first present the PINOCCHIO dark matter halo catalog that we will use to estimate parameter posteriors. Secondly, we perform cluster count cosmological analyses using our new hybrid likelihood, which we compare with standard methods.

\subsection{The PINOCCHIO dark matter halo catalogs}

We use one simulated dark matter halo catalog generated using the PINOCCHIO algorithm (PINpointing Orbit-Crossing Collapsed HIerarchical Objects, \citet{Monaco_2022,munari2017improving}). PINOCCHIO is based on Lagrangian Perturbation Theory \citep{moutarde1991precollapse,buchert1992lagrangian,bouchet1994perturbative} and ellipsoidal collapse \citep{bond1996peak,eisenstein1994analytical,Monaco1997}, it is known to reproduce the halo mass function, halo two-point statistics, and halo bias with an accuracy of $5-10\%$ across the relevant redshift range that we will consider in our analyses. PINOCCHIO has been used to generate a thousand dark matter halo catalogs at the best-fit cosmology from \citet{Planck2014cosmology}; the best fit cosmological parameters are $\Omega_m$ = 0.30711, $\Omega_b$ = 0.048254, $h$ = 0.6777, $n_s$ = 0.96, $\sigma_8$= 0.8288. Each simulation covers a quarter of the sky and encompasses a redshift range from 0 to 2.5 with virial halo masses greater than $2.45 \times 10^{13} h^{-1}M_\odot$. Then, the PINOCCHIO halo masses have been re-scaled to abundance-match the mass function averaged over 1,000 realizations with the \citet{Despali_2015} analytic fit. 

\subsection{Application to PINOCCHIO data}
We use the package {\tt PySSC} to compute the $S_{ij}$ matrix at the \textit{Planck} cosmology used in the PINOCCHIO algorithm. We consider the sky fraction $f_{\rm sky} = 1/4$ provided by PINOCCHIO. The effect of masking is difficult to implement in the prediction of $\sigma^2(z_1,z_2)$ \citep{Lacasa_2018}. To account for the limited sky fraction of the PINOCCHIO simulation, we re-scale the full sky $S_{ij}^{\rm fullsky}$ in \cref{eq:S_ij} such as $S_{ij}^{\rm partialsky} =S_{ij}^{\rm fullsky}/f_{\rm sky}$. While this is only an approximation, \citet{Gouyou2022SSC} showed that this approach performs quite well for sufficiently large sky areas. We keep the covariance of the matter density fluctuations $\{S_{ij}\}_{1\leq i,j\leq n_z}$ fixed at \textit{Planck} cosmology in the Bayesian inference since it is computationally demanding. Using a fiducial $S_{ij}$ does not induce noticeable differences in the results compared to using a cosmology-dependent covariance \citep{Fumagalli2021pinocchiovariance}. For the halo mass function and the halo bias, we use respectively the \citet{Despali_2015} and \citet{Tinker_2010} implementations available in the Core Cosmology Library (CCL, \citet{Chisari_2019}). The halo mass function, comoving volume, and halo bias are calculated at each position in the parameter space on a mass-redshift grid and interpolated. Then the different functions are evaluated at the position of each cluster.
In order to evaluate the posterior distribution using the likelihood in \cref{eq:hybrid_likelihood_unbinned}, we use a halo catalog chosen at random from the 1,000 PINOCCHIO mocks. 
\begin{figure*}
\centering\includegraphics[width=.49\textwidth]{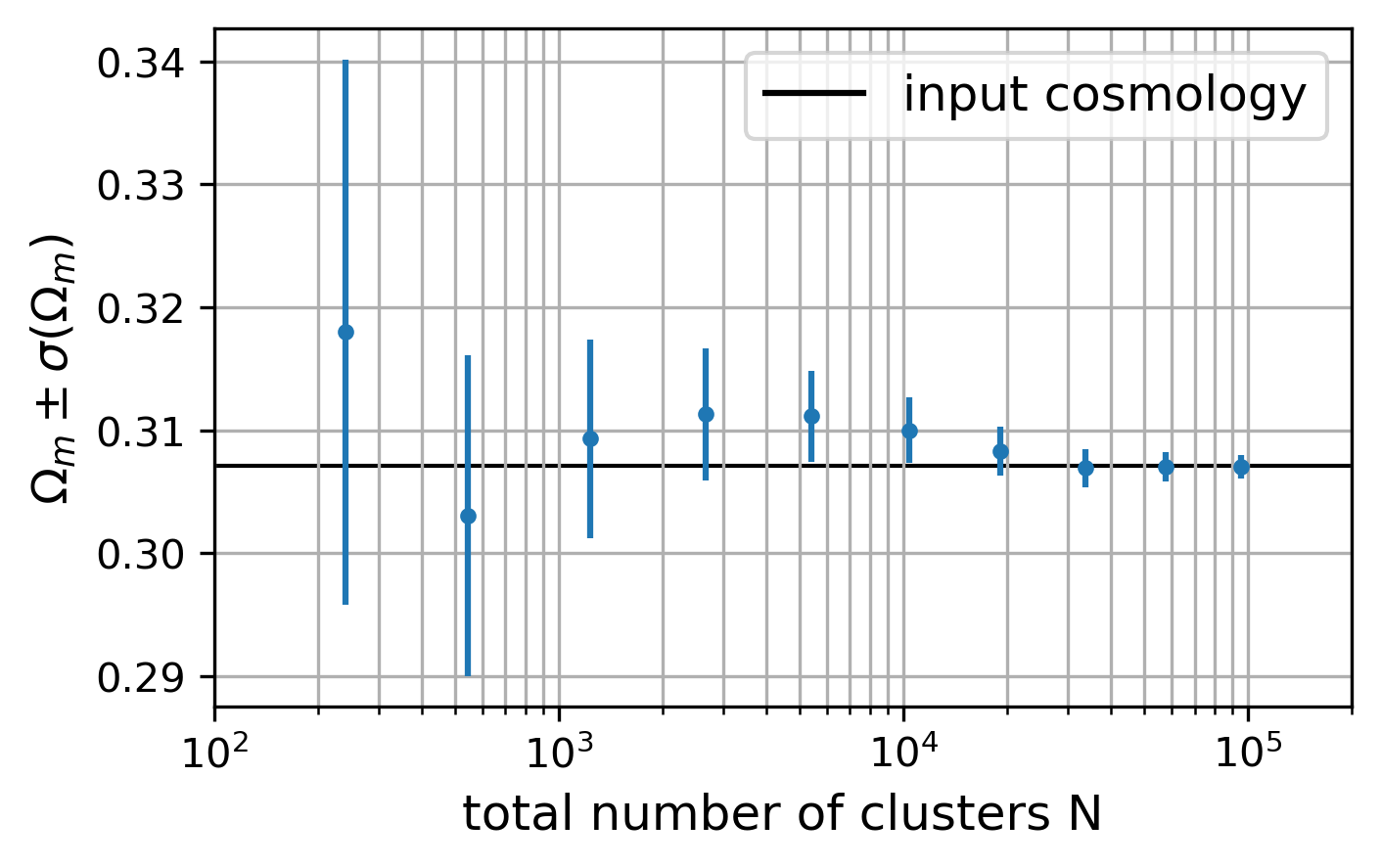}\includegraphics[width=.49\textwidth]{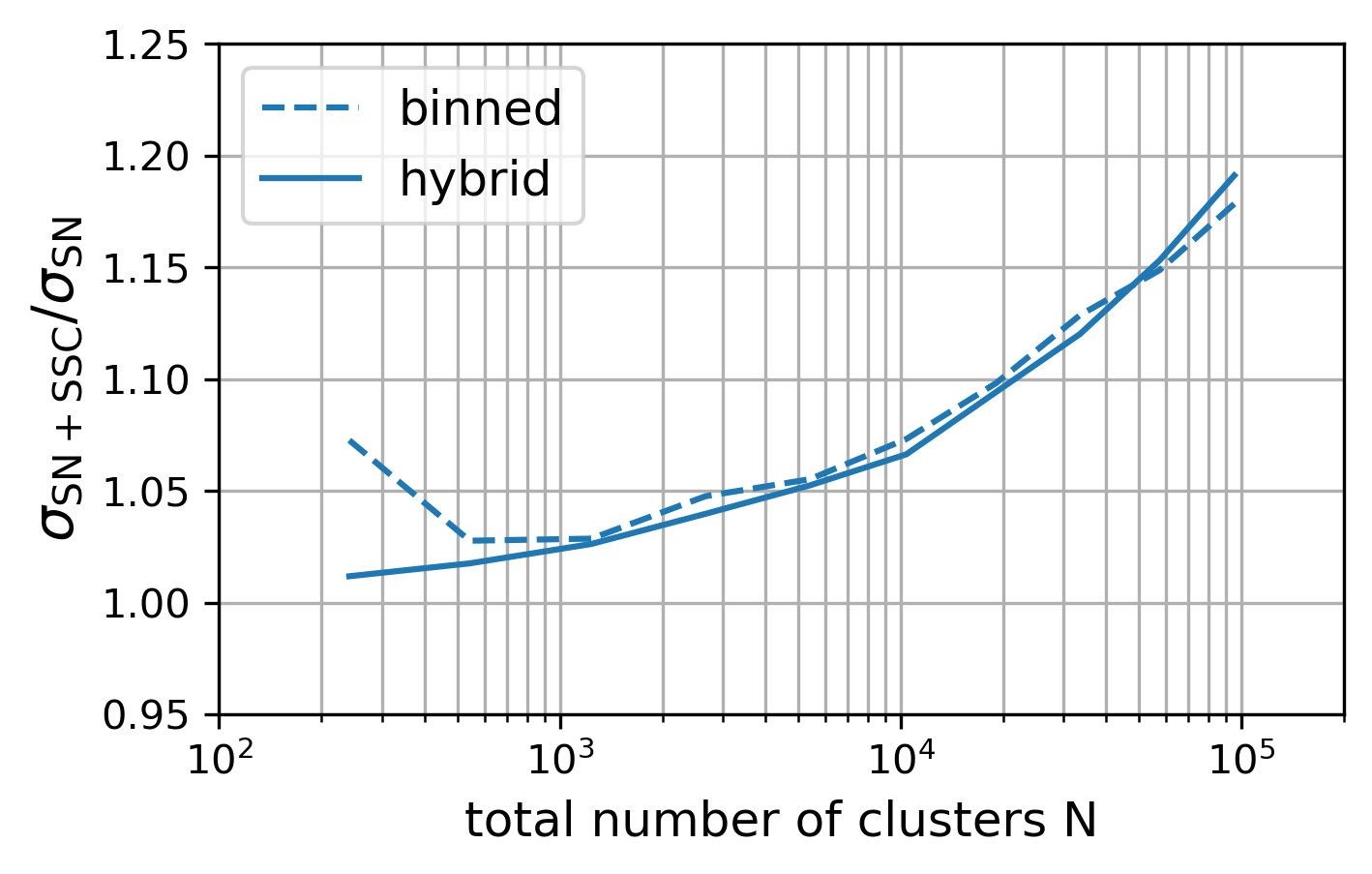}
    \caption{Left: Best fitted $\Omega_m$ and its dispersion for the different cluster samples presented in \cref{sec:pinochio} using the hybrid likelihood in \cref{eq:hybrid_likelihood_unbinned}.  The horizontal bar is the input PINOCCHIO cosmology. Right: Total parameter error divided by the Poisson-only error as a function of the cluster sample size. The dashed line represents the binned approach (for $\sigma_{\rm tot}$ and $\sigma_{\rm SN}$, we use respectively the Poisson likelihood in \cref{eq:poisson_distrib} and the Gaussian likelihood in \cref{eq:binned_gaussian_likelihood}) and the full line represents the hybrid approach.}
    \label{fig:res_Omegam}
\end{figure*}

Given a cluster sample and a cluster count likelihood, we reconstruct the parameter posterior for the parameter $\Omega_m$, namely $\mathcal{P}(\Omega_m|\widehat{d}) \propto \mathcal{L}(\widehat{d}|\Omega_m)\pi(\Omega_m)$ where $\widehat{d}$ denotes the cluster data. We tabulate the likelihood for different values of $\Omega_m$ (1,000 equally-spaced points within the range $[0.2, 0.6]$). We reconstruct the posterior within the flat-prior range $\Omega_m \in [0.2, 0.6]$, and normalize the product $\mathcal{L}(\widehat{d}|\Omega_m)\pi(\Omega_m)$ to 1. We then compute the posterior mean
\begin{equation}
    \langle \Omega_m\rangle = \int\ d\theta\ \Omega_m\ \mathcal{P}(\Omega_m|\widehat{d}),
\end{equation}
and the posterior variance
\begin{equation}
    \sigma^2(\Omega_m) = \int\ d\Omega_m\ (\Omega_m - \langle \Omega_m\rangle)^2\ \mathcal{P}(\Omega_m|\widehat{d}).
\end{equation} 
We first test whether the Poisson-only hybrid likelihood (i.e. by considering $\langle \mathcal{W}\rangle_{\rm SSC} = 1$ in \cref{eq:hybrid_likelihood_unbinned}) introduces some bias in the recovered $\Omega_m$ parameter, compared to the standard unbinned likelihood in \cref{eq:P_unbinned_Poisson}. From the selected random PINOCCHIO mock, we consider the cluster sample within the mass range $[\log_{10}M_{\rm min},\log_{10}M_{\rm min}] = [14.2, 15.6]$ and the redshift range $[z_{\rm min}, z_{\rm max}] = [0.2, 1.2]$. We first constrain $\Omega_m = 0.3074 \pm 0.0008$ using the standard unbinned likelihood in \cref{eq:P_unbinned_Poisson}, and second by using the Poisson-only hybrid likelihood by considering different numbers of redshift bins, from 1 (i.e. considering the full redshift range as a single bin) to 10 bins. \cref{fig:res_Omegam_posterior} (left panel) shows the bias in the recovered $\Omega_m$ parameter (blue crosses) and it dispersion $\sigma(\Omega_m)$ (orange dots) when using the Poisson-only hybrid likelihood compared to the standard unbinned one. We see that the mean and dispersion obtained from the Poisson-only hybrid likelihood are biased at most by 1$\%$ compared to the standard unbinned constraints, demonstrating that the hybrid formalism fairly compares to the standard unbinned description, providing the same constraints (mean and dispersion). 

Second, we investigate how the effect of SSC accounted for in the hybrid regime (trough the coefficient $\langle\mathcal{W}\rangle_{\rm SSC}$) impact the parameter constraints. We consider different cluster samples derived from the PINOCCHIO selected mock, by modifying the mass range within the redshift range $[0.2, 1.2]$. We consider $[\log_{10}M_{\rm min},15.6]$, where $\log_{10}M_{\rm min} = 14.2, 14.3, 14.4, 14.5, 14.6, 14.7, 14.8, 14.9, 15, 15.1$. For each configuration, we use the hybrid likelihood (Poisson-only or accounting for SSC) and we split the redshift range into three equal-sized bins to define the hybrid regime. We show in \cref{fig:res_Omegam_posterior} the posterior distribution of $\Omega_m$ for 5 different cluster samples whose total number of clusters range from $1,200$ to $960,000$ clusters. As the number of clusters increases, the dispersion of the posterior decreases, with and without SSC. The dashed lines correspond to the Poisson-only likelihood (i.e. by considering $\langle \mathcal{W}\rangle_{\rm SSC} = 1$ in \cref{eq:hybrid_likelihood_unbinned}), the full line corresponds to the hybrid likelihood in \cref{eq:hybrid_likelihood_unbinned}. 

Using the hybrid likelihood accounting for SSC in \cref{eq:hybrid_likelihood_unbinned}, the best fits and errors $\Omega_m \pm \sigma(\Omega_m)$ for the different cluster samples are plotted in \cref{fig:res_Omegam} (left panel) sorted by the total number of clusters in the sample. The error bar on the $\Omega_m$ parameter decreases with the cluster sample size, with roughly $1-2\sigma$ compatibility with the input cosmology (indicated by the black horizontal line). We have checked that it is a statistical fluctuation by redoing the analysis on another PINOCCHIO mock (over the 1,000 mocks, we recover on average the input cosmology, see \citet{Payerne2023}). Compared to the Poisson-only likelihood (i.e. by considering $\langle \mathcal{W}\rangle_{\rm SSC} = 1$ in \cref{eq:hybrid_likelihood_unbinned}), adding SSC slightly shifts\footnote{From \cref{eq:hybrid_Garrell}, the $\langle\mathcal{W} \rangle_{\rm SSC}$ is a convex function, with minimum reached at a given value $\Omega_m^{\rm min}$ roughly given by $\langle Nb \rangle_{\rm th}(\Omega_m^{\rm min}) = \langle Nb \rangle_{\rm obs}$. This minimum may slightly differ from $\Omega_m^{\rm min'}$ given by $\langle N\rangle_{\rm th}(\Omega_m^{\rm min'}) = \langle N\rangle_{\rm obs}$, since $\Omega_m^{\rm min}$ depends on the modeling of the halo bias. The difference between $\Omega_m^{\rm min}$ and $\Omega_m^{\rm min'}$ shifts slightly the parameter recovered from the hybrid+SSC likelihood compared to the Poisson-only hybrid likelihood. In this work, we use the \citet{Tinker_2010} halo bias, as used by \citet{Fumagalli2021pinocchiovariance,Payerne2023} who studied the PINOCCHIO mocks.} the best fit by at most 0.5 $\sigma$, but increases the parameter error bars, as expected for instance, from binned approaches explored in \citet{Fumagalli2021pinocchiovariance,Payerne2023}. The full line in \cref{fig:res_Omegam} (right panel) shows the ratio between the error $\sigma_{\rm tot}$ obtained with the hybrid likelihood accounting for SSC and $\sigma_{\rm SN}$, obtained with the Poisson-only likelihood (again, by considering $\langle \mathcal{W}\rangle_{\rm SSC} = 1$ in \cref{eq:hybrid_likelihood_unbinned}). We see that the ratio is always larger than one, as expected since SSC is an additive source of scatter in the count statistics, whatever the regime we consider. As a result, as we increase the sample to lower masses, the SSC contribution increases and impacts the errors by 20$\%$ when reaching $10^5$ clusters, whereas having no impact for small cluster samples with less than 1,000 clusters. 

For consistency, we also repeated the analysis using a binned approach. We consider 3 equal-sized redshift bins (as for the hybrid approach) and 4 log-spaced mass bins over each cluster sample mass interval. We consider the binned Poisson (BLC/Poisson) likelihood 
\begin{equation}
\label{eq:poisson_distrib}
\mathcal{L}^{\rm Poisson}_{\rm BLC} = \prod_{k=1}^{n_b} \frac{N_k^{\widehat{N}_k}}{\widehat{N}_k!}e^{-N_k}
\end{equation}
and the binned Gaussian likelihood (BLC/Gauss-SN+SSC), which is given by
\begin{equation}
    \mathcal{L}_{\rm BLC}^{\rm Gauss-SN+SSC} \propto |\Sigma|^{-1}\exp \left[ -\frac{1}{2}[N-\widehat{N}]^T\Sigma^{-1}[N-\widehat{N}] \right]
    \label{eq:binned_gaussian_likelihood}
\end{equation}
where $\Sigma$ is the binned count covariance, that is given by \citep{Lacasa2023SSC}
\begin{equation}
    \Sigma_{ij} = N\delta^K_{ij}+ N_iN_j\langle b\rangle_i\langle b\rangle_j S_{ij}, 
    \label{eq:covariance}
\end{equation}
where the first term in the above sum is the Poisson shot noise and the second is the binned SSC contribution. A detailed description of these two likelihoods can be found within \citet{Payerne2023}. The SN+SSC-to-SN ratio, i.e. between the total error (using the Gaussian binned likelihood in \cref{eq:binned_gaussian_likelihood}) and the Poisson-only error (using the Poisson binned likelihood in \cref{eq:poisson_distrib}) is shown in dashed lines in \cref{fig:res_Omegam} (right panel). We find very consistent results with the hybrid case, which validates that the SSC we have implemented in the hybrid formalism behaves as expected. Moreover, we have found that the hybrid errors are very similar to binned errors (at the percent level, whether considering SSC or not), denoting that the hybrid regime provides rather similar constraints on cosmological parameters compared to the binned approach since the halo mass function is rather monotonous in the log-mass space. However, as stated in the introduction, the unbinned regime (here hybrid) is useful for combining counts with multi-wavelength cluster observation, proxies, individual lensing profiles, etc., which is not explored in this work.  Again, the variability of the lines plotted in \cref{fig:res_Omegam} (right panel) is verified when considering another PINOCCHIO mock, we get the same result with some scatter, but without modifying our conclusions.

\section{Conclusions and discussions}
\label{sec:conclusions}
Inferring posteriors for cosmological parameters from observations, with a likelihood-based approach, necessitates a likelihood function that accurately describes the abundance of galaxy clusters. The unbinned formalism is often used to more easily include covariance between individual cluster properties. The standard unbinned method is based on Poisson statistics, in this paper we have presented a new likelihood to include the effect of super-sample covariance.
We found that incorporating SSC in the unbinned likelihood from the Taylor expansion of the GPC description requires several conditions, that are not fulfilled when redshift slices become small. The improved statistical description of the overdensity field \citep{Coles1991density,Wen2020unbinnedPNL} as well as incorporating truncated statistics (see e.g. \citet{Hu_2006}) are required to reach a reliable formula for the unbinned likelihood with SSC. To solve these issues by using the GPC description, we introduce a \textit{hybrid} likelihood, where the clusters are counted in standard redshift bins and infinitesimal mass bins. We have found a general form for the effect of super-sample covariance in the hybrid regime, showing that the hybrid Poisson-only likelihood needs to be corrected by an overall factor $\langle \mathcal{W}\rangle_{\rm SSC}$. We tested this new formalism using an idealistic dataset, the dark matter halo catalogs generated by the PINOCCHIO algorithm \citep{Monaco_2022}. 
We have found that the impact of SSC on parameter error is around $5\%$ for $<1,000$ cluster samples, but about 20$\%$ for cluster samples comprising tens of thousand clusters, indicating that SSC in the hybrid regime will have a strong impact on the cosmological parameter inference in the context of upcoming stage IV surveys, such as for Rubin LSST and \textit{Euclid}-like statistics.

In this analysis, we build the principal components for an unbiased estimation of cosmological parameters using cluster abundance in the hybrid regime. We tested the formalism in an idealistic framework with known masses and redshifts, but it should be supplemented with other sources of uncertainties in abundance statistics, such as the scatter of cluster proxies and reconstructed cluster masses \citep{Pratt2019impact,Salvati2020impact}, selection function \citep{Gallo2022completeness}, photometric redshifts but also the uncertainty in the calibration of the halo mass function \citep{Artis_2022}. The computational challenge arising when evaluating cosmological models in the unbinned regime when accounting for all these effects should be seriously addressed when forecasting the analysis of future large cluster samples, such as provided by the Rubin LSST of \textit{Euclid}.

Our new \textit{hybrid} likelihood is still an approximation of the true hybrid statistics. We have proposed in \cite{Payerne2023} a method to test the relevancy of a given cluster likelihood in the binned regime, namely the Poisson, the Gaussian-SSC, and the Gauss-Poisson Compound. By using the set of 1,000 PINOCCHIO dark matter halo catalogs (we only used one in this paper) and comparing the dispersion of posterior means to the individual posterior dispersion, we were able to test if a given likelihood was lacking SSC or not. We have found, in the binned regime, that the Gauss-SN+SSC and GPC likelihoods described well the full statistics of binned abundance, whereas the Poisson likelihood still misses SSC for Rubin LSST or \textit{Euclid}-like statistics. We aim to apply this methodology to test the robustness of our new \textit{hybrid} likelihood, for various sample sizes, from \textit{Planck}-like to \textit{Euclid}-like samples. This will enable us to know, in another manner, if the effect of SSC is modeled with sufficient precision in our new hybrid likelihood, since it is still an approximation of the true hybrid likelihood. 
\section*{ACKNOWLEDGEMENTS}
The authors thank the referee for their thorough and insightful remarks that have helped to significantly improve the article. The authors thank Pierluigi Monaco, Alex Saro, and Alessandra Fumagalli for providing the PINOCCHIO catalog used in this work. We thank Fabien Lacasa and Sylvain Gouyou-Beauchamps for useful discussions and for their help in using the PySSC code. We gratefully acknowledge support from the CNRS/IN2P3 Computing Center (Lyon - France) for providing computing and data-processing resources needed for this work.
\section*{DATA availability}
The PINOCCHIO simulations used in this article cannot be shared publicly. The data will be shared on reasonable request to Alessandra Fumagalli (alessandra.fumagalli@inaf.it). For reproducibility, we provide the codes used to combine {PySSC} and CCL, and to compute the hybrid likelihood. The codes are available at \href{https://github.com/payerne/LikelihoodsClusterAbundance}{https://github.com/payerne/LikelihoodsClusterAbundance}. 
\bibliographystyle{mnras}
\bibliography{main}
\appendix
\section{Unbinned approach with SSC used in Pacaud et al. (2018)}
\label{app:pacaud}
\input{appendix/appendix_pacaud}

\section{Hybrid likelihood with SSC}
\label{app:alternative_hybrid}
\input{appendix/appendix_Garrell}
\section{Calculation details of the Taylor expansion of the GPC likelihood}
\label{app:W2}
\input{appendix/appendix_W2}
\label{lastpage}
\end{document}

%% file: appendix/appendix_pacaud.tex
In this appendix, we review the methodology proposed by \citet{Mantz2010CCmethodunbinned} and \citet{Pacaud2018XXLCL} to include the effect of SSC in the unbinned description of cluster count. \citet{Mantz2010CCmethodunbinned} (see their Section 4.1.2) have proposed that the distribution of clusters in the mass and redshift space can be described by \textit{joint} likelihood of the total cluster count $\widehat{N}_{\rm obs}$ and observations $\mathcal{O}=\{m_k, z_k\}_{1 \leq k \leq N_{\rm obs}}$. Indeed, the unbinned likelihood is given by
\begin{equation}
    \mathcal{L}(\{\widehat{N}_k\}_{1 \leq k \leq n_b}) = \mathrm{e}^{-N_{\rm th}}\prod_{k = 1}^{\widehat{N}_{\rm tot}}n_h(m_k, z_k).
\end{equation}
By multiplying  and dividing each $n_h(m_k, z_k)$ by $N_{\rm th}$, we get that
\begin{equation}
     \mathcal{L}(\widehat{N}_{\rm obs}, \{m_k, z_k\}_{1 \leq k \leq \widehat{N}_{\rm obs}}) \propto \mathcal{P}(\widehat{N}_{\rm tot}|N_{\rm th})\prod_{k = 1}^{\widehat{N}_{\rm tot}}\frac{n_h(m_k, z_k)}{N_{\rm th}}.
\end{equation}
The left term $\mathcal{P}(\widehat{N}_{\rm tot}|N_{\rm th})$ is the Poisson likelihood for the total number of clusters, $N_{\rm th}$ being the predicted total number of clusters and the right term gives the likelihood of the \textit{detected} observable $\mathcal{O} = \{m_k, z_k\}_{1 \leq k \leq \widehat{N}_{\rm tot}}$. \citet{Pacaud2018XXLCL} have proposed to change the Poisson distribution $\mathcal{P}(\widehat{N}_{\rm tot}|N_{\rm th})$ with the single-variate GPC likelihood, so describing the total number of clusters as random variable scattered by both Poisson noise and SSC. So \citet{Pacaud2018XXLCL} have proposed an "unbinned" likelihood accounting for SSC, based on a methodology that is relatively different from what we have explored in this article, since:
\begin{enumerate}
    \item From the unbinned description of counts (see \citet{Cash1979unbinned}) we adopted, we have considered the joint likelihood of \textit{all} bins with observed cluster count $\widehat{N}_k = \{0,1\}$ (either they are empty or not) whereas \citet{Mantz2010CCmethodunbinned} have proposed an unbinned description of the observed clusters.
    \item We made use of a joint likelihood of $n_b$ dimensions (for the total number of bins), whereas \citet{Mantz2010CCmethodunbinned} used a likelihood with $\widehat{N}_{\rm obs} + 1$ dimension. We used the GPC formalism at the unbinned level, whereas \citet{Pacaud2018XXLCL} used the GPC formalism to describe the statistical properties of the total cluster count.
    \item The \textit{sample space} (space of possible outcomes) in the \citet{Mantz2010CCmethodunbinned} method (unbinned part) are the mass and redshift $m_k$ and $z_k$ for a given cluster in the sample. The per-cluster distribution $n_h(m_k, z_k)/N_{\rm th}$ is normalized over the mass and redshift ranges, i.e.
    \begin{equation}
        \int dz dm \frac{n_h(m, z)}{N_{\rm th}} = 1.
    \end{equation}
    In contrast, in the unbinned prescription from \citet{Cash1979unbinned} that we used in this paper, the outcomes are counts $N_k=\{0,1\}$ for a collection of small but \textit{fixed} mass-redshift bins. The unbinned prescription for counts is then normalized to all possible counts in the same mass-redshift bin, i.e. $p_0 + p_1 = 1$. The approach from \citet{Mantz2010CCmethodunbinned} may be, in some sense, associated with a Lagrangian interpretation of the cluster distribution (i.e. the position of each object in the mass-redshift space). In contrast, the \citet{Cash1979unbinned} description coincides with an Eulerian interpretation of the cluster population, counting them in fixed cells. 
\end{enumerate}

%% file: appendix/appendix_Garrell.tex
We propose to derive the hybrid unbinned likelihood including SSC from the methodology presented in \cite{Garrel2022XMM} (see their Appendix B). We start from the likelihood in \cref{eq:GPC_likelihood_multi}, and we get
\begin{align}
    \mathcal{W} &= \prod_k^{n_b} \exp\{-N_k b_k \delta_k\}(1 + b_k \delta_k)^{\widehat{N}_k}\\ 
    &=  \prod_k^{n_b} \exp\{-N_k b_k \delta_k+\widehat{N}_k\ln(1+b_k\delta_k)\}\\
    \label{eq:approx}
    &\approx \prod_k^{n_b} \exp\{-b_k\delta_k(N_k - \widehat{N}_k) - \frac{1}{2}\widehat{N}_kb_k^2\delta^2_k\}\\
    &=\prod_{m=1}^{n_z}\exp\{-\delta_m[\langle Nb \rangle_m - \langle \widehat{Nb} \rangle_m] - \delta_m^2\frac{1}{2}\langle \widehat{Nb^2} \rangle_m\} \\
    &=\exp\{v^T \Vec{\delta} - \Vec{\delta}^TC_1\Vec{\delta}\},
\end{align}
where $[C_1]_{mn} = \frac{1}{2}\mathrm{diag}(\langle\widehat{Nb^2} \rangle_n \delta_{mn}^K)$ and $v = [\langle Nb \rangle_n - \langle \widehat{Nb} \rangle_n]$. $\langle Nb \rangle_n$ is defined in \cref{eq:Nb_m_th}. The quantities $\langle \widehat{Nb} \rangle_n$ and $\langle \widehat{Nb^2} \rangle_n$ are properly defined respectively in \cref{eq:Nb_m_obs} and \cref{eq:N2b2_m_obs}, and can be derived similarly for a binned approach, since they are sums over the "observed" number of clusters $\widehat{N}_k$ in each bin, the limit to the unbinned regime is straightforward by turning $\widehat{N}_k$ to 0 or 1. The complete likelihood is given by averaging the above equation over all possible values of $\delta$. Using the multidimensional Gaussian integral as encountered in quantum field theory (see e.g. \citet{Zee2003,Zinn2002})
\begin{align}
\begin{split}
    \int_{\mathrm{R}^{n_z}} d\Vec{\delta}\ \exp\{-\frac{1}{2}\Vec{\delta}^TS^{-1}\Vec{\delta}&-\Vec{\delta}^TC_1\Vec{\delta} + v^T\Vec{\delta}\}\\ &= \sqrt{\frac{\pi^{n_z}}{\det C}}\exp\{\frac{1}{4}v^TC^{-1}v\},
    \label{eq:Gaussian_integral}
\end{split}
\end{align}
where $S$ is the covariance of matter density fluctuations in redshift-limited volumes and using 
\begin{equation}
    [C]_{mn} = \frac{1}{2}[S_{mn}^{-1} + \langle\widehat{Nb^2} \rangle_n \delta_{mn}^K],
\end{equation}
we get that
\begin{equation}
    \langle \mathcal{W}\rangle_{\rm SSC} = \sqrt{\frac{\det S^{-1}}{\det C}}\exp\{\frac{1}{4}v^TC^{-1}v\}.
\end{equation}
The approximate formula in \cref{eq:approx} is valid if, in each redshift bin we have
\begin{equation}
    \frac{1}{3}\sum_{k=1}^{\widehat{N}_{\rm tot}^{m}}\widehat{N}_k (b_k\delta_k)^3 \ll 1.
    \label{eq:approx_condition}
\end{equation}
Such a condition depends on the redshift range spanned by the cluster sample, the sky fraction, and the binning scheme of the redshift space for the hybrid prescription. Writing $\langle b_k \rangle\delta_k = \delta N_{\rm tot, loc}^m/N_{\rm tot}^{m}$ where $N_{\rm tot}^{m}$ is the total predicted cluster count in redshift bin $m$ and $\delta N_{\rm tot, loc}^m = (N_{\rm tot, loc}^{m} - N_{\rm tot}^{m})$ is the shift of the predicted count due to SSC, we get that \cref{eq:approx_condition} $\propto (\delta N_{\rm tot, loc}^m)^3/(N_{\rm tot}^{m})^2$. As explained in \citet{Garrel2022XMM}, for large surveys where $N_{\rm tot}^{m} \gg 1$, $\delta N_{\rm tot, loc}^m$ can be compared at most to the Poisson shot noise $\sqrt{N_{\rm tot}^{m}}$, such that the condition in \cref{eq:approx_condition} is verified.


%% file: appendix/appendix_W2.tex
In this appendix, we review the calculation of the abundance likelihood using the methodology presented in \citet{2014_Takada}. 

\begin{table*}
\centering
\begin{tabular}{c||c|c}
\hline
Coefficients&Binned analysis &Unbinned analysis\\
\hline
[1] &$\sum\limits_{m,n=1}^{n_z} S_{mn}\sum\limits_{i,j=1}^{n_m}b_{im}b_{jn}N_{im}N_{jn}$&$\bigints dz_1 dm_1 n_h(m_1,z_1)b(m_1, z_1)\bigints dz_2 dm_2\ n_h(m_2, z_2)b(m_2, z_2) \sigma^2(z_1, z_2)$\\

[2] &$\sum\limits_{m,n=1}^{n_z} S_{mn}\sum\limits_{i,j=1}^{n_m}b_{im}b_{jn}N_{im}\widehat{N}_{jn}$&$\sum\limits_{k = 1}^{\widehat{N}_{\rm tot}}b(m_k, z_k)\bigints dz_1 dm_1 n_h(m_1, z_1)b(m_1, z_1)\sigma^2(z_1, z_k)$\\

[3] &$\sum\limits_{m,n=1}^{n_z} S_{mn}\sum\limits_{i,j=1}^{n_m}b_{im}b_{jn}\widehat{N}_{im}\widehat{N}_{jn}$&$\sum\limits_{k,l = 1}^{\widehat{N}_{\rm tot}}b(m_k, z_k)b(m_l, z_l)\sigma^2(z_k, z_l)$\\

[4] &$\sum\limits_{m=1}^{n_z} S_{mm} \sum\limits_{i=1}^{n_m}b_{im}b_{im}\widehat{N}_{im}$&$\sum\limits_{k = 1}^{\widehat{N}_{\rm tot}}b(m_k, z_k)^2\sigma^2(z_k, z_k)$\\

\end{tabular}
\caption{Coefficients of the $\langle\mathcal{W}\rangle_{\rm SSC}  = 1+\frac{1}{2}([1] - 2\times [2] + [3]-[4])$ contribution in the binned (middle column) and unbinned regime (right column).}
\label{tab:1234}
\end{table*}

\subsection{Second order Taylor expansion}
\label{sec:W2_subsec}
We can construct the full likelihood of the cluster counts $\{\widehat{N}_k\}$ at fixed overdensity $\{\delta_k\}$ as given by the product of each Poisson likelihood, such as 
\begin{equation}
    \mathcal{L}(\{\widehat{N}_{i}\}|\{x_k\}) = \prod_{k=1}^{n_b}\mathcal{P}(\widehat{N}_k|N_k) \mathcal{W} = \mathcal{L}_{\rm BLC}^{\rm Poisson}\mathcal{W},
\label{eq:likelihood_fixed_overdensity}
\end{equation}
where $\mathcal{L}_{\rm BLC}^{\rm Poisson}$ is the standard binned Poisson likelihood. To compute the \textit{effective} likelihood in \cref{eq:GPC_likelihood_multi} in the low abundance regime, \citet{2014_Takada} considered two different assumptions. The first is $b_k N_k \delta_k  \ll 1$ denoting the small abundance regime, and the second is $b_k  \delta_k  \ll 1$, that is that the averaged density perturbation in a given redshift bin is very small (for sufficiently large redshift bins). Considering the Taylor expansion of each Poisson law $\mathcal{P}(\widehat{N}_k|x_k)$ in \cref{eq:Poisson_law} according to both conditions, we get
\begin{align}
    \mathcal{P}(\widehat{N}_k|x_k) &=\frac{1}{\widehat{N}_k!}e^{-N_k(1 + b_k\delta_k)}N_k^{\widehat{N}_k}(1 + b_k\delta_k)^{\widehat{N}_k}\\ &\approx (1 + w_k^{(1)}\delta_k + w_k^{(2)}\delta_k^2)\mathcal{P}(\widehat{N}_k|N_k),
    \label{eq:Poisson_k_xk}
\end{align}
where
\begin{align}
\label{eq:w1_k}
    w^{(1)}_k  &= b_k(\widehat{N}_k - N_k),\\
    \label{eq:w2_k}
    w^{(2)}_k &=  \frac{1}{2}b_k^2[(N_k - \widehat{N}_k)^2 - \widehat{N}_k].
\end{align}
We get that the extra term  $\mathcal{W}$ is given by
\begin{align}
    \mathcal{W} &\approx  \prod_{k=1}^{n_b}(1 + w_k^{(1)}\delta_k + w_k^{(2)}\delta_k^2) \\&= 1 + \sum_{k=1}^{n_b}\mathcal{W}_k^{(1)} \delta_k + \sum_{k,l=1}^{n_b}\mathcal{W}^{(2)}_{kl} \delta_{k}\delta_{l} + o(\delta_{k}^3),
    \label{eq:W_delta2_Taylor}
\end{align}
where
\begin{equation}
    \mathcal{W}_k^{(1)} = \frac{\partial \mathcal{W}}{\partial \delta_k}(\{\delta_i = 0\}) = w^{(1)}_k,
\end{equation}
and
\begin{equation}
    \mathcal{W}^{(2)}_{kl} = \frac{1}{2}\frac{\partial^2 \mathcal{W}}{\partial \delta_k\partial \delta_l}(\{\delta_i = 0\}) = \left\{
    \begin{array}{ll}
        w^{(2)}_k & \mbox{if } k=l, \\
        \frac{1}{2}w^{(1)}_kw^{(1)}_l & \mbox{else. }
    \end{array}
\right.
\end{equation}
After replacing the coefficients $w_k^{(1)}$ in \cref{eq:w1_k} and $w_k^{(2)}$ in \cref{eq:w2_k}, we get the result
\begin{align}
\begin{split}
    \mathcal{W} &= 1 + \sum_{k=1}^{n_b}\mathcal{W}_k^{(1)} \delta_k + \\&\frac{1}{2}\sum_{k,l=1}^{n_b}\delta_{k}\delta_{l} b_k b_l [(\widehat{N}_k - N_k) (\widehat{N}_l - N_l) - \delta_{kl}^K\widehat{N}_k],
    \label{eq:W_delta}
\end{split}
\end{align}
where $\delta_{kl}^K$ is the Kronecker delta.where $\mathcal{W}_k^{(1)}$ and $\mathcal{W}^{(2)}_{kl}$ are respectively the first and second order coefficient in the Taylor expansion of $\mathcal{W}$, both depending on the coefficients $w_k^{(1)}$ and $w_k^{(2)}$ (see calculation details in \cref{app:W2}). We define the GPC likelihood given by the average of the local likelihood in \cref{eq:likelihood_fixed_overdensity} over all possible values of the fields $\delta_k$ in each redshift bin. We get 
\begin{equation}
    \mathcal{L}(\{\widehat{N}_{i}\}) =  \mathcal{L}_{\rm BLC}^{\rm Poisson}\langle \mathcal{W}\rangle_{\rm SSC}\,,
    \label{eq:L_GPC_binned_approx}
\end{equation}
where
\begin{align}
    \langle \mathcal{W}\rangle_{\rm SSC} &\approx 1 + \frac{1}{2}\sum_{k,l=1}^{n_b}S_{kl} b_k b_l [(\widehat{N}_k - N_k) (\widehat{N}_l - N_l) - \delta_{kl}^K\widehat{N}_k]\\&= 1+\frac{1}{2}([1] - 2\times [2] + [3]-[4]).
    \label{eq:W_average}
\end{align}
and $\delta_{kl}^K$ is the Kronecker delta function. The coefficients [1], [2], [3], and [4] are given in \cref{tab:1234} (second column), where we add specific indexes for both mass and redshift bins such that $[k]$ that previously denoted the $k$-th mass-redshift bins are now given by two indexes $[ik]$ such as the indexes $ik$ refer to the $i$-th mass and $k$-th redshift bins. So the total number of bins $n$ can be written as $n_b = n_z\times n_m$, where $n_z$ is the total number of redshift bins, and $n_m$ is the total number of mass bins. 
The average product between two counts $i$ and $j$ is given by $\langle \widehat{N}_i\widehat{N}_j\rangle = N_i \delta^K_{ij} + N_iN_j(1 + b_ib_j S_{ij})$, where $\langle\cdot \rangle$ denotes the average over the likelihood distribution in \cref{eq:L_GPC_binned_approx}. 
Using that $\langle \widehat{N}_i\rangle = N_i$, the variance of a single cluster count is given by $\sigma^2(\widehat{N}_i) = N_i + N_i^2 b_i^2 S_{ii}$, and 
the covariance between two different mass-redshift bins writes $\mathrm{Cov}(\widehat{N}_i, \widehat{N}_j) = N_i N_j b_i b_j S_{ij}$, so GPC describes the intrinsic count variance arising from Poisson sampling as well as SSC. This result is consistent with the binned approach using a Gausian+SSC (see e.g., \citet{Abbott2020DESCL})
or a non-approximated GPC likelihood (see e.g., \citet{Lesci2022KIDSCL}), where the cluster count variance has an extra-term accounting for SSC. To check our results, we consider all clusters to be in a single redshift bin, and we show that we obtain the same result as \citet{2014_Takada}. We consider the same matter density fluctuation amplitude $S_0$ in a narrow redshift bin, setting $S_{kl} = S_0$.  We obtain, (in agreement with their Eq. (8))

\begin{equation}
    \langle \mathcal{W}\rangle_{\rm SSC}= 1+\frac{1}{2}\left[\left(\sum_{k = 1}^{n_b} b_k(N_k - \widehat{N}_k)\right)^2 - \sum_{k = 1}^{n_b} b_k^2\widehat{N}_k\right]S_0.
\end{equation}
\citet{Hu_2006} also derived this formula for a single mass and redshift bin (see their Eq. (15)). So, from \cref{eq:W_average}, we have included several redshift bins and correlations between them. We note that in the binned framework $b_k \delta_k \ll 1$ is satisfied as the density fluctuations averaged over a sufficiently large region will be very small, however, it is not simple to show that this is necessarily true as we move to smaller bins and therefore smaller volumes.

Finally, assuming that the description of SSC is valid through the $S_{ij}$ terms in \cref{eq:S_ij} for any size of redshift interval, and using that the covariance of the matter density field between two narrow redshift intervals becomes
\begin{equation}
     \lim_{V_k, V_l \rightarrow dV_k dV_l} S_{kl} = \sigma^2(z_k, z_l),
     \label{eq:S_to_sigma}
\end{equation} 
the unbinned likelihood with SSC is given by (i) an unbinned Poisson contribution in \cref{eq:P_unbinned_Poisson} and (ii) by the new $\langle \mathcal{W}\rangle_{\rm SSC}$ term, whose coefficients $[1,2,3,4]$ are listed in \cref{tab:1234} (third column), after turning sums to integrals.

\begin{table*}
    \centering
    \begin{tabular}{c||l|l}
    \hline
    [i]&Expression&Derivative 
    $\mathrm{[i]}_{,n}(\{\delta_i = 0\})$\\
    \hline
    I&$\sum\limits_{a = 1}^{n_b}(f_a)_{,klm}\prod\limits_{b\neq a}^{n_b} f_b $&$\sum\limits_{a = 1}^{n_b}(f_a)_{,klmn}+\sum\limits_{a = 1}^{n_b}\sum\limits_{b \neq a}^{n_b}(f_a)_{,klm}(f_b)_{,n}$\\
    \hline
    II&$\sum\limits_{a = 1}^{n_b}\sum\limits_{b\neq a}^{n_b}(f_a)_{,kl}(f_b)_{,m}\prod\limits_{c\neq (a,b)}^{n_b} f_c$&$\sum\limits_{a = 1}^{n_b}\sum\limits_{b\neq a}^{n_b}\left((f_a)_{,kln}(f_b)_{,m} + (f_a)_{,kl}(f_b)_{,mn} + (f_a)_{,kl}(f_b)_{,m}\sum\limits_{c\neq(a,b)}^{n_b}(f_c)_{,n}\right)$\\
    \hline
    III&$\sum\limits_{a = 1}^{n_b}\sum\limits_{b\neq a}^{n_b}(f_a)_{,km}(f_b)_{,l}\prod\limits_{c\neq (a,b)}^{n_b} f_c$&$\sum\limits_{a = 1}^{n_b}\sum\limits_{b\neq a}^{n_b}\left((f_a)_{,kmn}(f_b)_{,l} + (f_a)_{,km}(f_b)_{,ln} + (f_a)_{,km}(f_b)_{,l}\sum\limits_{c\neq(a,b)}^{n_b}(f_c)_{,n}\right)$\\
    \hline
    IV&$\sum\limits_{a = 1}^{n_b}\sum\limits_{b\neq a}^{n_b}(f_a)_{,k}(f_b)_{,lm}\prod\limits_{c\neq (a,b)}^{n_b}f_c$&$\sum\limits_{a = 1}^{n_b}\sum\limits_{b\neq a}^{n_b}\left((f_a)_{,kn}(f_b)_{,lm} + (f_a)_{,k}(f_b)_{,lmn} + (f_a)_{,k}(f_b)_{,lm}\sum\limits_{c\neq(a,b)}^{n_b}(f_c)_{,n}\right)$\\
    \hline
    V&$\sum\limits_{a = 1}^{n_b}\sum\limits_{b\neq a}^{n_b}\sum\limits_{c\neq (a,b)}^{n_b}(f_a)_{,k}(f_b)_{,l} (f_c)_{,m}\prod\limits_{d\neq(a,b,c) }^{n_b}f_d$&$\sum\limits_{a = 1}^{n_b}\sum\limits_{b\neq a}^{n_b}\sum\limits_{c\neq (a,b)}^{n_b}\Biggl((f_a)_{,kn}(f_b)_{,l} (f_c)_{,m}+(f_a)_{,k}(f_b)_{,ln} (f_c)_{,m}\Biggr)+$\\
    &&$\sum\limits_{a = 1}^{n_b}\sum\limits_{b\neq a}^{n_b}\sum\limits_{c\neq (a,b)}^{n_b}\left((f_a)_{,k}(f_b)_{,l} (f_c)_{,mn}+(f_a)_{,k}(f_b)_{,l} (f_c)_{,m}\sum\limits_{d\neq(a,b,c)}^{n_b}(f_d)_{,n}\right)$\\
    \end{tabular}
    \caption{Second column: coefficients [I-II-III-IV-V] of the $W_{,klm}$ term. Third column: Derivatives of [I-II-III-IV-V] with respect to $\delta_n$, evaluated at $\{\delta_i = 0\}$ for $i\in [1, n]$.}
    \label{tab:coeff_W}
\end{table*}

\begin{enumerate}
    \item As proposed in \citet{2014_Takada}, we used the conditions $b_k N_k \delta_k  \ll 1$ and $b_k  \delta_k  \ll 1$ to perform the Taylor expansion. These two conditions are valid for relatively \textit{small} mass bins and for sufficiently \textit{large} redshift bins, such that $S_{ii}$ becomes sufficiently small; From \cref{fig:sigma_Sigma}, the amplitude of the $S_{ii}$ terms are smaller as we increase the size of redshift bins, for a given redshift range. However, for thinner redshift bins, (e.g. for the unbinned regime), $b_k  \delta_k  \ll 1$ will not be valid. Releasing $b_k  \delta_k  \ll 1$ may lead to a change in the results we obtained. For large $\delta_k$, the Taylor expansion of the $(1 + b_k\delta_k)^{\widehat{N}_k}$ term at second order may not be sufficient in \cref{eq:Poisson_k_xk}. To address this issue, we can choose that the unbinned observations are $\widehat{N}_k = \{0,1\}$ in \cref{eq:Poisson_k_xk} leaving only two choices, $1$ or $(1 + b_k\delta_k)$, namely $(1 + b_k\widehat{N}_k\delta_k)$. Then, we use the $N_k\delta_k b_k \ll 1$ limit in the $\exp{(-N_kb_k\delta_k)}$ terms. We recalculate the $w_k^{(1,2)}$ coefficients, and we find that $w_k^{(1)}$ is unchanged but $w_k^{(2)} = \frac{1}{2}b_k(N_k -\widehat{N}_k)^2 - \widehat{N}_k^2$ (the last term has changed from $ - \widehat{N}_k$ to $ - \widehat{N}_k^2$). We then repeat Taylor expansion to the second order in $\delta_k$ in \cref{eq:W_delta2_Taylor}. 
As a result, we found that the last term in the sum in \cref{eq:W_average} that was given by $ - \widehat{N}_k\delta^K_{kl} $ is now given by $ - \widehat{N}_k^2\delta^K_{kl}$. In the unbinned regime, $\widehat{N}_k = \widehat{N}_k^2$. so we have verified that the form of the likelihood is still valid. 

\item When considering infinitesimal redshift bins, the variance of the matter density fluctuations is maximal and larger by one or two orders in magnitude compared to binned approaches. In this case, the average of the Poisson distributions over $\delta_k$ in \cref{eq:GPC_likelihood_multi} may lead to nonphysical Poisson means, i.e. $(1 + b_k\delta_k) < 0$. Then, the individual Poisson distributions are not properly defined, and the GPC description fails to describe binned abundance. It is even larger when $f_{\rm sky} < 1$ (since $\sigma_{\rm partialsky}$ can be approximated to $\sigma_{\rm fullsky}/\sqrt{f_{\rm sky}}$, see e.g. \citet{Lacasa_2018}). For instance, with $f_{\rm sky} = 1/4$, $b\sigma_{\rm partialsky}$ reaches $\sim$ 0.3. So the approximation $b_k\delta_k \ll 1$ only stands for small values of $\sigma_{\rm partialsky}(z,z)b(m,z)$, that depends on the mass-redshift range probed by the cluster sample and the sky area. Let us note that this problem can be solved using log-normal probabilities. Instead of the prescription $x_k = N_k(1 + b_k\delta_k)$, we first define the random variable $y_k = \ln x_k$ that follows a Gaussian distribution. Hence, $x_k$ follows a log-normal distribution and is always positive. This formulation is however difficult to handle in practice since it requires the derivation of the covariance matrix for this new log-normal field $y_k$. We did not explore this in our work. 

\item We used the Taylor series to the second order in $\delta_k$ in \cref{eq:W_delta2_Taylor} to reach the above equation. The next non-zero term in the Taylor expansion is the fourth given by,

\begin{equation}
\frac{1}{4!}\sum_{k,l,m,n}^{n_b}\frac{\partial^4\mathcal{W}}{\partial \delta_k\partial\delta_l\partial\delta_m\partial\delta_n}\langle \delta_k\delta_l\delta_m\delta_n\rangle,
\end{equation}

\noindent The amplitude $\langle \delta_k\delta_l\delta_m\delta_n\rangle$ of this term is smaller, proportional to $S_{kl}^2$. We do not account for the third-order correlation functions, since $\langle \delta_k \delta_l \delta_m \rangle = 0$ for any Gaussian random fields. Calculating higher-order derivatives of $\mathcal{W}$ analytically proves very difficult, which we explore in the next section.
\end{enumerate}
\subsection{Fourth-order Taylor expansion}
\label{sec:W4_subsec}
Here we explore how to derive the $\mathcal{W}_{,klmn}$ coefficients, that appear in the Taylor expansion of $\mathcal{W}$. For that, we use the notation
\begin{equation}
    \mathcal{W} = \prod_{a=1}^{n_b} f_a,
\end{equation}
where the indexes $a,b,c,d$ refer to each function $f_a$, and $k,l,m,n$ refer to the variables $\delta_k$ and corresponding derivatives. So we get the first-order derivative
\begin{equation}
    \mathcal{W}_{,k} = \sum_{a = 1}^{n_b}(f_a)_{,k}\prod_{b\neq a}^{n_b} f_b,
\end{equation}
where $(f_a)_{,k} = \partial f_a/\partial \delta_k$. The second-order derivative is
\begin{align}
    \mathcal{W}_{,kl} = \sum_{a = 1}^{n_b}(f_a)_{,kl}\prod_{b\neq a} f_b+ \sum_{a = 1}^{n_b}(f_a)_{,k}\sum_{b\neq a}^{n_b}(f_b)_{,l}\prod_{c\neq (a,b)} f_c.
\end{align}
The above equation was used to derive the \cref{eq:W_average}, i.e. up to the second order in the Taylor expansion. The third-order derivative is given by
\begin{align}
  \mathcal{W}_{,klm}  = \mathrm{[I]+[II]+[III]+[IV]+[V]}.
\end{align}
The coefficients $\mathrm{[i]}$ are listed in \cref{tab:coeff_W} (second column). The $\mathcal{W}_{,klmn}^{(i)}$ are given by
\begin{align}
    \mathcal{W}_{,klmn} &= \mathrm{[I]}_{,n}+\mathrm{[II]}_{,n}+\mathrm{[III]}_{,n}+\mathrm{[IV]}_{,n}+\mathrm{[V]}_{,n}\\&=\mathcal{W}_{,klmn}^{(1)}+\mathcal{W}_{,klmn}^{(2)}+\mathcal{W}_{,klmn}^{(3)}+\mathcal{W}_{,klmn}^{(4)}
\end{align}
The coefficients $\mathrm{[i]}_{,n}$ are listed in \cref{tab:coeff_W} (third column), and $\mathcal{W}_{,klmn}$ is given by the sum of each line. We can re-arrange $\mathcal{W}_{,klmn}$ defining the $\mathcal{W}_{,klmn}^{(1,2,3,4)}$ coefficients, respectively defined by the sum of the product of 1, 2, 3 or 4 functions $f_a$ (with a total of four derivatives shared by the corresponding functions). We get
\begin{equation}
    \mathcal{W}_{,klmn}^{(1)}= \sum\limits_{a = 1}^{n_b}(f_a)_{,klmn},
\end{equation}

\begin{align}
\begin{split}
\mathcal{W}_{,klmn}^{(2)}=&\sum\limits_{a = 1}^{n_b}\sum\limits_{b\neq a}^{n_b}\Biggl((f_a)_{,klm}(f_b)_{,n}+(f_a)_{,kln}(f_b)_{,m} \\& +(f_a)_{,kl}(f_b)_{,mn}+(f_a)_{,kmn}(f_b)_{,l} \\&+ (f_a)_{,km}(f_b)_{,ln}+(f_a)_{,kn}(f_b)_{,lm} \\&+ (f_a)_{,k}(f_b)_{,lmn}\Biggr),
\end{split}
\end{align}

\begin{align}
\begin{split}
\mathcal{W}_{,klmn}^{(3)}=&\sum\limits_{a = 1}^{n_b}\sum\limits_{b\neq a}^{n_b}\sum\limits_{c\neq (a,b)}^{n_b}\Biggl((f_a)_{,kl}(f_b)_{,m}(f_c)_{,n} \\&+ (f_a)_{,km}(f_b)_{,l}(f_c)_{,n} + (f_a)_{,k}(f_b)_{,lm}(f_c)_{,n}\\&+(f_a)_{,kn}(f_b)_{,l}(f_c)_{,m} +(f_a)_{,k}(f_b)_{,ln}(f_c)_{,m}\\&+(f_a)_{,k}(f_b)_{,l}(f_c)_{,mn}\Biggr),
\end{split}
\end{align}

\begin{equation}
\mathcal{W}_{,klmn}^{(4)}=\sum\limits_{a = 1}^{n_b}\sum\limits_{b\neq a}^{n_b}\sum\limits_{c\neq (a,b)}^{n_b}\sum\limits_{d\neq (a,b,c)}^{n_b}(f_a)_{,k}(f_b)_{,l} (f_c)_{,m}(f_d)_{,n}.
\end{equation}
We then write down the explicit values of $(f_a)_{,k...n}$ from the Poisson distribution in \cref{eq:Poisson_k_xk}, where each $f_a$ is the individual probability excess due to SSC for the bin with index $a$ and is given by \begin{equation}
    f_a = 1 + w_a^{(1)}\delta_a+w_a^{(1)}\delta_a+w_a^{(2)}\delta_a^2+w_a^{(3)}\delta_a^3+w_a^{(4)}\delta_a^4
\end{equation}
we get the first derivative $(f_a)_{,k} = \delta^K_{ka} w^{(1)}_a$, the second order derivative 
    $(f_a)_{,kl} = 2\delta^K_{kl}\delta^K_{ka} w^{(2)}_a$, the third order $(f_a)_{,klm} = 6\delta^K_{ka}\delta^K_{la}\delta^K_{ma} w^{(3)}_a$
    and the fourth order $(f_a)_{,klmn}= 24\delta^K_{ka}\delta^K_{la}\delta^K_{ma}\delta^K_{na}w^{(4)}_a$.
The $w_a^{(1)}$ and $w_a^{(2)}$ are given in \cref{eq:w1_k} and \cref{eq:w2_k}. For the unbinned prescription, we get $(1 + b_k\delta_k)^{\widehat{N}_k} = (1 + \widehat{N}_kb_k\delta_k)$, we get that
\begin{align}
    w^{(3)}_a &= \frac{b_k^3\widehat{N}_kN_k^2}{2!} - \frac{b_k^3N^3_k}{3!},\\
    w^{(4)}_a &= -\frac{b_k^4\widehat{N}_kN_k^3}{3!}+\frac{N_k^4b_k^4}{4!}.
\end{align}
From these above equations, we see that obtaining a usable expression for the fourth-order correction of the likelihood is not straightforward. In effect, one could benefit from the closing relation of the Kronecker Delta function given by $\sum\limits_{i=1}^{n_b}\delta^K_{ij}g_i = g_j$ to write synthetically $\mathcal{W}_{,klmn}$ but here the sums do not run on all indices, but on restricted portions of the indices (e.g. for $c\neq (a,b)$). Once accomplished by using the relevant combinatory method, we need to use Wick's probability theorem \citep{Wick1950} for a Gaussian random field, given by $\langle\delta_k \delta_l\delta_m\delta_n\rangle = S_{kl}S_{mn} + S_{km}S_{ln} + S_{kn}S_{lm}$. We can use the combination and symmetry properties of the above equations to obtain the explicit form of $\mathcal{W}$ up to the fourth order in terms of $w^{(1,2,3,4)}_k$. For instance, The simplest coefficient $\mathcal{W}_{,klmn}^{(1)}$ is given by $\mathcal{W}_{,klmn}^{(1)} = 24\delta^K_{kl}\delta^K_{lm}\delta^K_{kn} w^{(4)}_k$.

%% file: main.bbl
\begin{thebibliography}{}
\makeatletter
\relax
\def\mn@urlcharsother{\let\do\@makeother \do\$\do\&\do\#\do\^\do\_\do\%\do\~}
\def\mn@doi{\begingroup\mn@urlcharsother \@ifnextchar [ {\mn@doi@}
  {\mn@doi@[]}}
\def\mn@doi@[#1]#2{\def\@tempa{#1}\ifx\@tempa\@empty \href
  {http://dx.doi.org/#2} {doi:#2}\else \href {http://dx.doi.org/#2} {#1}\fi
  \endgroup}
\def\mn@eprint#1#2{\mn@eprint@#1:#2::\@nil}
\def\mn@eprint@arXiv#1{\href {http://arxiv.org/abs/#1} {{\tt arXiv:#1}}}
\def\mn@eprint@dblp#1{\href {http://dblp.uni-trier.de/rec/bibtex/#1.xml}
  {dblp:#1}}
\def\mn@eprint@#1:#2:#3:#4\@nil{\def\@tempa {#1}\def\@tempb {#2}\def\@tempc
  {#3}\ifx \@tempc \@empty \let \@tempc \@tempb \let \@tempb \@tempa \fi \ifx
  \@tempb \@empty \def\@tempb {arXiv}\fi \@ifundefined
  {mn@eprint@\@tempb}{\@tempb:\@tempc}{\expandafter \expandafter \csname
  mn@eprint@\@tempb\endcsname \expandafter{\@tempc}}}

\bibitem[\protect\citeauthoryear{{Abbott} et~al.,}{{Abbott}
  et~al.}{2020}]{Abbott2020DESCL}
{Abbott} T.,  et~al., 2020, \mn@doi [\prd] {10.1103/PhysRevD.102.023509}, \href
  {https://ui.adsabs.harvard.edu/abs/2020PhRvD.102b3509A} {102, 023509}

\bibitem[\protect\citeauthoryear{{Abdullah}, {Klypin}  \& {Wilson}}{{Abdullah}
  et~al.}{2020}]{Abdullah2020SDSSCL}
{Abdullah} M.~H.,  {Klypin} A.,   {Wilson} G.,  2020, \mn@doi [\apj]
  {10.3847/1538-4357/aba619}, \href
  {https://ui.adsabs.harvard.edu/abs/2020ApJ...901...90A} {901, 90}

\bibitem[\protect\citeauthoryear{Ade et~al.,}{Ade
  et~al.}{2014}]{Ade2014PlanckCL}
Ade P.,  et~al., 2014, \mn@doi [\aap] {10.1051/0004-6361/201321521}, 571, A20

\bibitem[\protect\citeauthoryear{Ade et~al.,}{Ade
  et~al.}{2016}]{Ade2016PlanckCL}
Ade P.,  et~al., 2016, \mn@doi [\aap] {10.1051/0004-6361/201525833}, 594, A24

\bibitem[\protect\citeauthoryear{Artis, Melin, Bartlett  \& and}{Artis
  et~al.}{2022}]{Artis_2022}
Artis E.,  Melin J.-B.,  Bartlett J.,   and C.~M.,  2022, \mn@doi [{EPJ} Web of
  Conferences] {10.1051/epjconf/202225700004}, 257, 00004

\bibitem[\protect\citeauthoryear{{Aymerich} et~al.,}{{Aymerich}
  et~al.}{2024}]{Aymerich2024PlanckChandraCL}
{Aymerich} G.,  et~al., 2024, \mn@doi [arXiv e-prints]
  {10.48550/arXiv.2402.04006}, \href
  {https://ui.adsabs.harvard.edu/abs/2024arXiv240204006A} {p. arXiv:2402.04006}

\bibitem[\protect\citeauthoryear{{Bartlett}}{{Bartlett}}{1997}]{Bartlett1997}
{Bartlett} J.~G.,  1997, \mn@doi [Astronomical Society of the Pacific
  Conference Series] {10.48550/arXiv.astro-ph/9703090}, \href
  {https://ui.adsabs.harvard.edu/abs/1997ASPC..126..365B} {126, 365}

\bibitem[\protect\citeauthoryear{{Bocquet} et~al.,}{{Bocquet}
  et~al.}{2015}]{Bocquet2015SPTCL}
{Bocquet} S.,  et~al., 2015, \mn@doi [\apj] {10.1088/0004-637X/799/2/214},
  \href {https://ui.adsabs.harvard.edu/abs/2015ApJ...799..214B} {799, 214}

\bibitem[\protect\citeauthoryear{{Bocquet} et~al.,}{{Bocquet}
  et~al.}{2019}]{Bocquet2019SPTCL}
{Bocquet} S.,  et~al., 2019, \mn@doi [\apj] {10.3847/1538-4357/ab1f10}, \href
  {https://ui.adsabs.harvard.edu/abs/2019ApJ...878...55B} {878, 55}

\bibitem[\protect\citeauthoryear{{Bocquet} et~al.,}{{Bocquet}
  et~al.}{2023}]{Bocquet2023SPTDES}
{Bocquet} S.,  et~al., 2023, \mn@doi [arXiv e-prints]
  {10.48550/arXiv.2310.12213}, \href
  {https://ui.adsabs.harvard.edu/abs/2023arXiv231012213B} {p. arXiv:2310.12213}

\bibitem[\protect\citeauthoryear{{Bocquet} et~al.,}{{Bocquet}
  et~al.}{2024}]{Bocquet2024SPT}
{Bocquet} S.,  et~al., 2024, \mn@doi [arXiv e-prints]
  {10.48550/arXiv.2401.02075}, \href
  {https://ui.adsabs.harvard.edu/abs/2024arXiv240102075B} {p. arXiv:2401.02075}

\bibitem[\protect\citeauthoryear{{Bond} \& {Myers}}{{Bond} \&
  {Myers}}{1996}]{bond1996peak}
{Bond} J.~R.,  {Myers} S.~T.,  1996, \apjs

\bibitem[\protect\citeauthoryear{{Bouchet}, {Colombi}, {Hivon}  \&
  {Juszkiewicz}}{{Bouchet} et~al.}{1995}]{bouchet1994perturbative}
{Bouchet} F.~R.,  {Colombi} S.,  {Hivon} E.,   {Juszkiewicz} R.,  1995, \aap,
  \href {https://ui.adsabs.harvard.edu/abs/1995A&A...296..575B} {296, 575}

\bibitem[\protect\citeauthoryear{{Buchert}}{{Buchert}}{1992}]{buchert1992lagrangian}
{Buchert} T.,  1992, \mn@doi [\mnras] {10.1093/mnras/254.4.729}, \href
  {https://ui.adsabs.harvard.edu/abs/1992MNRAS.254..729B} {254, 729}

\bibitem[\protect\citeauthoryear{{Carlstrom} et~al.,}{{Carlstrom}
  et~al.}{2011}]{Carlstrom2011SPTwp}
{Carlstrom} J.~E.,  et~al., 2011, \mn@doi [\pasp] {10.1086/659879}, \href
  {https://ui.adsabs.harvard.edu/abs/2011PASP..123..568C} {123, 568}

\bibitem[\protect\citeauthoryear{{Cash}}{{Cash}}{1979}]{Cash1979unbinned}
{Cash} W.,  1979, \mn@doi [\apj] {10.1086/156922}, \href
  {https://ui.adsabs.harvard.edu/abs/1979ApJ...228..939C} {228, 939}

\bibitem[\protect\citeauthoryear{{Chaubal} et~al.,}{{Chaubal}
  et~al.}{2022}]{Chaubal2022SPTCLCMBlensing}
{Chaubal} P.~S.,  et~al., 2022, \mn@doi [\apj] {10.3847/1538-4357/ac6a55},
  \href {https://ui.adsabs.harvard.edu/abs/2022ApJ...931..139C} {931, 139}

\bibitem[\protect\citeauthoryear{{Chisari} et~al.,}{{Chisari}
  et~al.}{2019}]{Chisari_2019}
{Chisari} N.~E.,  et~al., 2019, \mn@doi [\apjs] {10.3847/1538-4365/ab1658},
  \href {https://ui.adsabs.harvard.edu/abs/2019ApJS..242....2C} {242, 2}

\bibitem[\protect\citeauthoryear{{Chiu}, {Klein}, {Mohr}  \& {Bocquet}}{{Chiu}
  et~al.}{2023}]{Chiu2022erositaCL}
{Chiu} I.~N.,  {Klein} M.,  {Mohr} J.,   {Bocquet} S.,  2023, \mn@doi [\mnras]
  {10.1093/mnras/stad957}, \href
  {https://ui.adsabs.harvard.edu/abs/2023MNRAS.522.1601C} {522, 1601}

\bibitem[\protect\citeauthoryear{{Cole} \& {Kaiser}}{{Cole} \&
  {Kaiser}}{1989}]{Cole1989clusering}
{Cole} S.,  {Kaiser} N.,  1989, \mn@doi [\mnras] {10.1093/mnras/237.4.1127},
  \href {https://ui.adsabs.harvard.edu/abs/1989MNRAS.237.1127C} {237, 1127}

\bibitem[\protect\citeauthoryear{{Coles} \& {Jones}}{{Coles} \&
  {Jones}}{1991}]{Coles1991density}
{Coles} P.,  {Jones} B.,  1991, \mn@doi [\mnras] {10.1093/mnras/248.1.1}, \href
  {https://ui.adsabs.harvard.edu/abs/1991MNRAS.248....1C} {248, 1}

\bibitem[\protect\citeauthoryear{Costanzi et~al.,}{Costanzi
  et~al.}{2019}]{Costanzi2019SDSSCL}
Costanzi M.,  et~al., 2019, \mn@doi [\mnras] {10.1093/mnras/stz1949}, 488, 4779

\bibitem[\protect\citeauthoryear{{Costanzi} et~al.,}{{Costanzi}
  et~al.}{2021}]{Costanzi2021DESSPTCL}
{Costanzi} M.,  et~al., 2021, \mn@doi [\prd] {10.1103/PhysRevD.103.043522},
  \href {https://ui.adsabs.harvard.edu/abs/2021PhRvD.103d3522C} {103, 043522}

\bibitem[\protect\citeauthoryear{{Despali}, {Giocoli}, {Angulo}, {Tormen},
  {Sheth}, {Baso}  \& {Moscardini}}{{Despali} et~al.}{2016}]{Despali_2015}
{Despali} G.,  {Giocoli} C.,  {Angulo} R.~E.,  {Tormen} G.,  {Sheth} R.~K.,
  {Baso} G.,   {Moscardini} L.,  2016, \mn@doi [\mnras]
  {10.1093/mnras/stv2842}, \href
  {https://ui.adsabs.harvard.edu/abs/2016MNRAS.456.2486D} {456, 2486}

\bibitem[\protect\citeauthoryear{{Eisenstein} \& {Loeb}}{{Eisenstein} \&
  {Loeb}}{1995}]{eisenstein1994analytical}
{Eisenstein} D.~J.,  {Loeb} A.,  1995, \mn@doi [\apj] {10.1086/175193}, \href
  {https://ui.adsabs.harvard.edu/abs/1995ApJ...439..520E} {439, 520}

\bibitem[\protect\citeauthoryear{{Ettori}, {Donnarumma}, {Pointecouteau},
  {Reiprich}, {Giodini}, {Lovisari}  \& {Schmidt}}{{Ettori}
  et~al.}{2013}]{Ettori2013Xmass}
{Ettori} S.,  {Donnarumma} A.,  {Pointecouteau} E.,  {Reiprich} T.~H.,
  {Giodini} S.,  {Lovisari} L.,   {Schmidt} R.~W.,  2013, \mn@doi [\ssr]
  {10.1007/s11214-013-9976-7}, \href
  {https://ui.adsabs.harvard.edu/abs/2013SSRv..177..119E} {177, 119}

\bibitem[\protect\citeauthoryear{{Evrard}, {Arnault}, {Huterer}  \&
  {Farahi}}{{Evrard} et~al.}{2014}]{Evrard2014multi}
{Evrard} A.~E.,  {Arnault} P.,  {Huterer} D.,   {Farahi} A.,  2014, \mn@doi
  [\mnras] {10.1093/mnras/stu784}, \href
  {https://ui.adsabs.harvard.edu/abs/2014MNRAS.441.3562E} {441, 3562}

\bibitem[\protect\citeauthoryear{{Fowler} et~al.,}{{Fowler}
  et~al.}{2007}]{Fowler2007actwp}
{Fowler} J.~W.,  et~al., 2007, \mn@doi [\ao] {10.1364/AO.46.003444}, \href
  {https://ui.adsabs.harvard.edu/abs/2007ApOpt..46.3444F} {46, 3444}

\bibitem[\protect\citeauthoryear{{Fumagalli} et~al.,}{{Fumagalli}
  et~al.}{2021}]{Fumagalli2021pinocchiovariance}
{Fumagalli} A.,  et~al., 2021, \mn@doi [\aap] {10.1051/0004-6361/202140592},
  \href {https://ui.adsabs.harvard.edu/abs/2021A&A...652A..21F} {652, A21}

\bibitem[\protect\citeauthoryear{{Fumagalli}, {Costanzi}, {Saro}, {Castro}  \&
  {Borgani}}{{Fumagalli} et~al.}{2023}]{Fumagalli2023SDSS}
{Fumagalli} A.,  {Costanzi} M.,  {Saro} A.,  {Castro} T.,   {Borgani} S.,
  2023, \mn@doi [arXiv e-prints] {10.48550/arXiv.2310.09146}, \href
  {https://ui.adsabs.harvard.edu/abs/2023arXiv231009146F} {p. arXiv:2310.09146}

\bibitem[\protect\citeauthoryear{{Gallo}, {Douspis}, {Soubri{\'e}}  \&
  {Salvati}}{{Gallo} et~al.}{2023}]{Gallo2022completeness}
{Gallo} S.,  {Douspis} M.,  {Soubri{\'e}} E.,   {Salvati} L.,  2023, \mn@doi
  [arXiv e-prints] {10.48550/arXiv.2309.11544}, \href
  {https://ui.adsabs.harvard.edu/abs/2023arXiv230911544G} {p. arXiv:2309.11544}

\bibitem[\protect\citeauthoryear{{Garrel} et~al.,}{{Garrel}
  et~al.}{2022}]{Garrel2022XMM}
{Garrel} C.,  et~al., 2022, \mn@doi [\aap] {10.1051/0004-6361/202141204}, \href
  {https://ui.adsabs.harvard.edu/abs/2022A&A...663A...3G} {663, A3}

\bibitem[\protect\citeauthoryear{{Ghirardini} et~al.,}{{Ghirardini}
  et~al.}{2024}]{Ghirardini2024erositaCL}
{Ghirardini} V.,  et~al., 2024, arXiv e-prints, \href
  {https://ui.adsabs.harvard.edu/abs/2024arXiv240208458G} {p. arXiv:2402.08458}

\bibitem[\protect\citeauthoryear{{Gouyou Beauchamps}, {Lacasa}, {Tutusaus},
  {Aubert}, {Baratta}, {Gorce}  \& {Sakr}}{{Gouyou Beauchamps}
  et~al.}{2022}]{Gouyou2022SSC}
{Gouyou Beauchamps} S.,  {Lacasa} F.,  {Tutusaus} I.,  {Aubert} M.,  {Baratta}
  P.,  {Gorce} A.,   {Sakr} Z.,  2022, \mn@doi [\aap]
  {10.1051/0004-6361/202142052}, \href
  {https://ui.adsabs.harvard.edu/abs/2022A&A...659A.128G} {659, A128}

\bibitem[\protect\citeauthoryear{{Hasselfield} et~al.,}{{Hasselfield}
  et~al.}{2013}]{Hasselfield2013ACTCL}
{Hasselfield} M.,  et~al., 2013, \jcap, \href
  {https://ui.adsabs.harvard.edu/abs/2013JCAP...07..008H} {2013, 008}

\bibitem[\protect\citeauthoryear{Hu \& Cohn}{Hu \& Cohn}{2006}]{Hu_2006}
Hu W.,  Cohn J.~D.,  2006, \mn@doi [Physical Review D]
  {10.1103/physrevd.73.067301}, 73

\bibitem[\protect\citeauthoryear{{Hu} \& {Kravtsov}}{{Hu} \&
  {Kravtsov}}{2003}]{2003_SV_HU}
{Hu} W.,  {Kravtsov} A.~V.,  2003, \mn@doi [\apj] {10.1086/345846}, \href
  {https://ui.adsabs.harvard.edu/abs/2003ApJ...584..702H} {584, 702}

\bibitem[\protect\citeauthoryear{{LSST Science Collaboration} et~al.,}{{LSST
  Science Collaboration} et~al.}{2009}]{LSST}
{LSST Science Collaboration} et~al., 2009, \mn@doi [arXiv e-prints]
  {10.48550/arXiv.0912.0201}, \href
  {https://ui.adsabs.harvard.edu/abs/2009arXiv0912.0201L} {p. arXiv:0912.0201}

\bibitem[\protect\citeauthoryear{{Lacasa} \& {Grain}}{{Lacasa} \&
  {Grain}}{2019}]{Lacasa19}
{Lacasa} F.,  {Grain} J.,  2019, \mn@doi [\aap] {10.1051/0004-6361/201834343},
  \href {https://ui.adsabs.harvard.edu/abs/2019A&A...624A..61L} {624, A61}

\bibitem[\protect\citeauthoryear{{Lacasa}, {Lima}  \& {Aguena}}{{Lacasa}
  et~al.}{2018}]{Lacasa_2018}
{Lacasa} F.,  {Lima} M.,   {Aguena} M.,  2018, \mn@doi [\aap]
  {10.1051/0004-6361/201630281}, \href
  {https://ui.adsabs.harvard.edu/abs/2018A&A...611A..83L} {611, A83}

\bibitem[\protect\citeauthoryear{{Lacasa} et~al.,}{{Lacasa}
  et~al.}{2023}]{Lacasa2023SSC}
{Lacasa} F.,  et~al., 2023, \mn@doi [\aap] {10.1051/0004-6361/202245148}, \href
  {https://ui.adsabs.harvard.edu/abs/2023A&A...671A.115L} {671, A115}

\bibitem[\protect\citeauthoryear{{Laureijs} et~al.,}{{Laureijs}
  et~al.}{2011}]{laureijs2011euclid}
{Laureijs} R.,  et~al., 2011, \mn@doi [arXiv e-prints]
  {10.48550/arXiv.1110.3193}, \href
  {https://ui.adsabs.harvard.edu/abs/2011arXiv1110.3193L} {p. arXiv:1110.3193}

\bibitem[\protect\citeauthoryear{{Lee} et~al.,}{{Lee}
  et~al.}{2019}]{S02019whitepaper}
{Lee} A.,  et~al., 2019, in Bulletin of the American Astronomical Society.
  p.~147 (\mn@eprint {arXiv} {1907.08284}), \mn@doi{10.48550/arXiv.1907.08284}

\bibitem[\protect\citeauthoryear{{Lee}, {Battye}  \& {Bolliet}}{{Lee}
  et~al.}{2024}]{Lee2024PlanckACTCL}
{Lee} E.,  {Battye} R.,   {Bolliet} B.,  2024, \mn@doi [arXiv e-prints]
  {10.48550/arXiv.2403.19542}, \href
  {https://ui.adsabs.harvard.edu/abs/2024arXiv240319542L} {p. arXiv:2403.19542}

\bibitem[\protect\citeauthoryear{Lesci et~al.,}{Lesci
  et~al.}{2022}]{Lesci2022KIDSCL}
Lesci G.~F.,  et~al., 2022, \mn@doi [\aap] {10.1051/0004-6361/202040194}, 659,
  A88

\bibitem[\protect\citeauthoryear{{Lima} \& {Hu}}{{Lima} \&
  {Hu}}{2004}]{Lima2024GPC}
{Lima} M.,  {Hu} W.,  2004, \mn@doi [\prd] {10.1103/PhysRevD.70.043504}, \href
  {https://ui.adsabs.harvard.edu/abs/2004PhRvD..70d3504L} {70, 043504}

\bibitem[\protect\citeauthoryear{{Mahdavi}, {Hoekstra}, {Babul}, {Bildfell},
  {Jeltema}  \& {Henry}}{{Mahdavi} et~al.}{2013}]{Mahdavi2013XMM}
{Mahdavi} A.,  {Hoekstra} H.,  {Babul} A.,  {Bildfell} C.,  {Jeltema} T.,
  {Henry} J.~P.,  2013, \mn@doi [\apj] {10.1088/0004-637X/767/2/116}, \href
  {https://ui.adsabs.harvard.edu/abs/2013ApJ...767..116M} {767, 116}

\bibitem[\protect\citeauthoryear{{Mana}, {Giannantonio}, {Weller}, {Hoyle},
  {H{\"u}tsi}  \& {Sartoris}}{{Mana} et~al.}{2013}]{Mana2013SDSScosmologyCL}
{Mana} A.,  {Giannantonio} T.,  {Weller} J.,  {Hoyle} B.,  {H{\"u}tsi} G.,
  {Sartoris} B.,  2013, \mn@doi [\mnras] {10.1093/mnras/stt1062}, \href
  {https://ui.adsabs.harvard.edu/abs/2013MNRAS.434..684M} {434, 684}

\bibitem[\protect\citeauthoryear{{Mantz}, {Allen}, {Rapetti}  \&
  {Ebeling}}{{Mantz} et~al.}{2010}]{Mantz2010CCmethodunbinned}
{Mantz} A.,  {Allen} S.~W.,  {Rapetti} D.,   {Ebeling} H.,  2010, \mn@doi
  [\mnras] {10.1111/j.1365-2966.2010.16992.x}, \href
  {https://ui.adsabs.harvard.edu/abs/2010MNRAS.406.1759M} {406, 1759}

\bibitem[\protect\citeauthoryear{{Mantz} et~al.,}{{Mantz}
  et~al.}{2015}]{Mantz2014WTGCL}
{Mantz} A.~B.,  et~al., 2015, \mn@doi [\mnras] {10.1093/mnras/stu2096}, \href
  {https://ui.adsabs.harvard.edu/abs/2015MNRAS.446.2205M} {446, 2205}

\bibitem[\protect\citeauthoryear{{Merloni} et~al.,}{{Merloni}
  et~al.}{2012}]{Merloni2012erositaWP}
{Merloni} A.,  et~al., 2012, \mn@doi [arXiv e-prints]
  {10.48550/arXiv.1209.3114}, \href
  {https://ui.adsabs.harvard.edu/abs/2012arXiv1209.3114M} {p. arXiv:1209.3114}

\bibitem[\protect\citeauthoryear{{Mo} \& {White}}{{Mo} \&
  {White}}{1996}]{Mo1996clustering}
{Mo} H.~J.,  {White} S.~D.~M.,  1996, \mn@doi [\mnras]
  {10.1093/mnras/282.2.347}, \href
  {https://ui.adsabs.harvard.edu/abs/1996MNRAS.282..347M} {282, 347}

\bibitem[\protect\citeauthoryear{{Monaco}}{{Monaco}}{1997}]{Monaco1997}
{Monaco} P.,  1997, \mn@doi [\mnras] {10.1093/mnras/287.4.753}, 287, 753

\bibitem[\protect\citeauthoryear{{Monaco}, {Theuns}  \& {Taffoni}}{{Monaco}
  et~al.}{2002}]{Monaco_2022}
{Monaco} P.,  {Theuns} T.,   {Taffoni} G.,  2002, \mn@doi [\mnras]
  {10.1046/j.1365-8711.2002.05162.x}, \href
  {https://ui.adsabs.harvard.edu/abs/2002MNRAS.331..587M} {331, 587}

\bibitem[\protect\citeauthoryear{{Moutarde}, {Alimi}, {Bouchet}, {Pellat}  \&
  {Ramani}}{{Moutarde} et~al.}{1991}]{moutarde1991precollapse}
{Moutarde} F.,  {Alimi} J.~M.,  {Bouchet} F.~R.,  {Pellat} R.,   {Ramani} A.,
  1991, \mn@doi [\apj] {10.1086/170728}, 382, 377

\bibitem[\protect\citeauthoryear{{Mu{\~n}oz-Echeverr{\'\i}a}
  et~al.,}{{Mu{\~n}oz-Echeverr{\'\i}a} et~al.}{2023}]{Echeverria2023SZmass}
{Mu{\~n}oz-Echeverr{\'\i}a} M.,  et~al., 2023, \mn@doi [\aap]
  {10.1051/0004-6361/202244981}, \href
  {https://ui.adsabs.harvard.edu/abs/2023A&A...671A..28M} {671, A28}

\bibitem[\protect\citeauthoryear{{Munari}, {Monaco}, {Sefusatti}, {Castorina},
  {Mohammad}, {Anselmi}  \& {Borgani}}{{Munari}
  et~al.}{2017}]{munari2017improving}
{Munari} E.,  {Monaco} P.,  {Sefusatti} E.,  {Castorina} E.,  {Mohammad} F.~G.,
   {Anselmi} S.,   {Borgani} S.,  2017, \mn@doi [\mnras]
  {10.1093/mnras/stw3085}, 465, 4658

\bibitem[\protect\citeauthoryear{{Murray}, {Bartlett}, {Artis}  \&
  {Melin}}{{Murray} et~al.}{2022}]{Murray2022lensing}
{Murray} C.,  {Bartlett} J.~G.,  {Artis} E.,   {Melin} J.-B.,  2022, \mn@doi
  [\mnras] {10.1093/mnras/stac689}, \href
  {https://ui.adsabs.harvard.edu/abs/2022MNRAS.512.4785M} {512, 4785}

\bibitem[\protect\citeauthoryear{Pacaud et~al.,}{Pacaud
  et~al.}{2018}]{Pacaud2018XXLCL}
Pacaud F.,  et~al., 2018, \mn@doi [\aap] {10.1051/0004-6361/201834022}, 620,
  A10

\bibitem[\protect\citeauthoryear{{Park}, {Sunayama}, {Takada}, {Kobayashi},
  {Miyatake}, {More}, {Nishimichi}  \& {Sugiyama}}{{Park}
  et~al.}{2023}]{Park2023lensingabundance}
{Park} Y.,  {Sunayama} T.,  {Takada} M.,  {Kobayashi} Y.,  {Miyatake} H.,
  {More} S.,  {Nishimichi} T.,   {Sugiyama} S.,  2023, \mn@doi [\mnras]
  {10.1093/mnras/stac3410}, \href
  {https://ui.adsabs.harvard.edu/abs/2023MNRAS.518.5171P} {518, 5171}

\bibitem[\protect\citeauthoryear{{Payerne}, {Murray}, {Combet}, {Doux},
  {Fumagalli}  \& {Penna-Lima}}{{Payerne} et~al.}{2023}]{Payerne2023}
{Payerne} C.,  {Murray} C.,  {Combet} C.,  {Doux} C.,  {Fumagalli} A.,
  {Penna-Lima} M.,  2023, \mn@doi [\mnras] {10.1093/mnras/stad489}, \href
  {https://ui.adsabs.harvard.edu/abs/2023MNRAS.520.6223P} {520, 6223}

\bibitem[\protect\citeauthoryear{Penna-Lima}{Penna-Lima}{2010}]{PennaLima_thesis}
Penna-Lima M.,  2010, Ph.D. Thesis, Centro Brasileiro de Pesquisas Físicas

\bibitem[\protect\citeauthoryear{{Penna-Lima}, {Makler}  \&
  {Wuensche}}{{Penna-Lima} et~al.}{2014}]{2014_PennaLima}
{Penna-Lima} M.,  {Makler} M.,   {Wuensche} C.~A.,  2014, \mn@doi [\jcap]
  {10.1088/1475-7516/2014/05/039}, \href
  {https://ui.adsabs.harvard.edu/abs/2014JCAP...05..039P} {2014, 039}

\bibitem[\protect\citeauthoryear{{Philcox}, {Spergel}  \&
  {Villaescusa-Navarro}}{{Philcox} et~al.}{2020}]{Philcox2020haloexclusion}
{Philcox} O. H.~E.,  {Spergel} D.~N.,   {Villaescusa-Navarro} F.,  2020,
  \mn@doi [\prd] {10.1103/PhysRevD.101.123520}, \href
  {https://ui.adsabs.harvard.edu/abs/2020PhRvD.101l3520P} {101, 123520}

\bibitem[\protect\citeauthoryear{{Pierre} et~al.,}{{Pierre}
  et~al.}{2016}]{xxl2016wp}
{Pierre} M.,  et~al., 2016, \mn@doi [\aap] {10.1051/0004-6361/201526766}, \href
  {https://ui.adsabs.harvard.edu/abs/2016A&A...592A...1P} {592, A1}

\bibitem[\protect\citeauthoryear{{Planck Collaboration} et~al.,}{{Planck
  Collaboration} et~al.}{2014}]{Planck2014cosmology}
{Planck Collaboration} et~al., 2014, \mn@doi [\aap]
  {10.1051/0004-6361/201321591}, \href
  {https://ui.adsabs.harvard.edu/abs/2014A&A...571A..16P} {571, A16}

\bibitem[\protect\citeauthoryear{{Poisson}}{{Poisson}}{1837}]{Poisson1837}
{Poisson} S.~D.,  1837, Bachelier, Imprimeur-Librairie

\bibitem[\protect\citeauthoryear{{Pratt}, {Arnaud}, {Biviano}, {Eckert},
  {Ettori}, {Nagai}, {Okabe}  \& {Reiprich}}{{Pratt}
  et~al.}{2019}]{Pratt2019impact}
{Pratt} G.~W.,  {Arnaud} M.,  {Biviano} A.,  {Eckert} D.,  {Ettori} S.,
  {Nagai} D.,  {Okabe} N.,   {Reiprich} T.~H.,  2019, \mn@doi [\ssr]
  {10.1007/s11214-019-0591-0}, \href
  {https://ui.adsabs.harvard.edu/abs/2019SSRv..215...25P} {215, 25}

\bibitem[\protect\citeauthoryear{{Rozo} et~al.,}{{Rozo}
  et~al.}{2010}]{Rozo2010CLSDSS}
{Rozo} E.,  et~al., 2010, \mn@doi [\apj] {10.1088/0004-637X/708/1/645}, \href
  {https://ui.adsabs.harvard.edu/abs/2010ApJ...708..645R} {708, 645}

\bibitem[\protect\citeauthoryear{{Salvati}, {Douspis}  \& {Aghanim}}{{Salvati}
  et~al.}{2018}]{Salvati2018Planck}
{Salvati} L.,  {Douspis} M.,   {Aghanim} N.,  2018, \mn@doi [\aap]
  {10.1051/0004-6361/201731990}, \href
  {https://ui.adsabs.harvard.edu/abs/2018A&A...614A..13S} {614, A13}

\bibitem[\protect\citeauthoryear{{Salvati}, {Douspis}  \& {Aghanim}}{{Salvati}
  et~al.}{2020}]{Salvati2020impact}
{Salvati} L.,  {Douspis} M.,   {Aghanim} N.,  2020, \mn@doi [\aap]
  {10.1051/0004-6361/202038465}, \href
  {https://ui.adsabs.harvard.edu/abs/2020A&A...643A..20S} {643, A20}

\bibitem[\protect\citeauthoryear{{Salvati} et~al.,}{{Salvati}
  et~al.}{2022}]{Salvati2022PlanckSPT}
{Salvati} L.,  et~al., 2022, \mn@doi [\apj] {10.3847/1538-4357/ac7ab4}, \href
  {https://ui.adsabs.harvard.edu/abs/2022ApJ...934..129S} {934, 129}

\bibitem[\protect\citeauthoryear{Sehgal et~al.,}{Sehgal
  et~al.}{2011}]{Sehgal2011actCL}
Sehgal N.,  et~al., 2011, \mn@doi [\apj] {10.1088/0004-637x/732/1/44}, 732, 44

\bibitem[\protect\citeauthoryear{{Smith} \& {Marian}}{{Smith} \&
  {Marian}}{2011}]{Smith2011GPC}
{Smith} R.~E.,  {Marian} L.,  2011, \mn@doi [\mnras]
  {10.1111/j.1365-2966.2011.19525.x}, \href
  {https://ui.adsabs.harvard.edu/abs/2011MNRAS.418..729S} {418, 729}

\bibitem[\protect\citeauthoryear{{Sunayama} et~al.,}{{Sunayama}
  et~al.}{2023}]{Sunayama2023HSClensing}
{Sunayama} T.,  et~al., 2023, \mn@doi [arXiv e-prints]
  {10.48550/arXiv.2309.13025}, \href
  {https://ui.adsabs.harvard.edu/abs/2023arXiv230913025S} {p. arXiv:2309.13025}

\bibitem[\protect\citeauthoryear{{Takada} \& {Spergel}}{{Takada} \&
  {Spergel}}{2014}]{2014_Takada}
{Takada} M.,  {Spergel} D.~N.,  2014, \mn@doi [\mnras] {10.1093/mnras/stu759},
  \href {https://ui.adsabs.harvard.edu/abs/2014MNRAS.441.2456T} {441, 2456}

\bibitem[\protect\citeauthoryear{{Tauber} et~al.,}{{Tauber}
  et~al.}{2010}]{Tauber2010planckwp}
{Tauber} J.~A.,  et~al., 2010, \mn@doi [\aap] {10.1051/0004-6361/200912983},
  \href {https://ui.adsabs.harvard.edu/abs/2010A&A...520A...1T} {520, A1}

\bibitem[\protect\citeauthoryear{{The Dark Energy Survey Collaboration}}{{The
  Dark Energy Survey Collaboration}}{2005}]{DES2005wpaper}
{The Dark Energy Survey Collaboration} 2005, \mn@doi [arXiv e-prints]
  {10.48550/arXiv.astro-ph/0510346}, \href
  {https://ui.adsabs.harvard.edu/abs/2005astro.ph.10346T} {pp
  astro--ph/0510346}

\bibitem[\protect\citeauthoryear{Tinker, Robertson, Kravtsov, Klypin, Warren,
  Yepes  \& Gottlöber}{Tinker et~al.}{2010}]{Tinker_2010}
Tinker J.~L.,  Robertson B.~E.,  Kravtsov A.~V.,  Klypin A.,  Warren M.~S.,
  Yepes G.,   Gottlöber S.,  2010, \mn@doi [The Astrophysical Journal]
  {10.1088/0004-637x/724/2/878}, 724, 878

\bibitem[\protect\citeauthoryear{To et~al.,}{To et~al.}{2021}]{To2021DEScomb}
To C.,  et~al., 2021, \mn@doi [Phys. Rev. Lett.]
  {10.1103/PhysRevLett.126.141301}, 126, 141301

\bibitem[\protect\citeauthoryear{{Truemper}}{{Truemper}}{1993}]{Truemper1993rosatwp}
{Truemper} J.,  1993, \mn@doi [Science] {10.1126/science.260.5115.1769}, \href
  {https://ui.adsabs.harvard.edu/abs/1993Sci...260.1769T} {260, 1769}

\bibitem[\protect\citeauthoryear{{Wen}, {Kemball}  \& {Saslaw}}{{Wen}
  et~al.}{2020}]{Wen2020unbinnedPNL}
{Wen} D.,  {Kemball} A.~J.,   {Saslaw} W.~C.,  2020, \mn@doi [\apj]
  {10.3847/1538-4357/ab6d6f}, \href
  {https://ui.adsabs.harvard.edu/abs/2020ApJ...890..160W} {890, 160}

\bibitem[\protect\citeauthoryear{Wick}{Wick}{1950}]{Wick1950}
Wick G.~C.,  1950, \mn@doi [Phys. Rev.] {10.1103/PhysRev.80.268}, 80, 268

\bibitem[\protect\citeauthoryear{{York} et~al.,}{{York}
  et~al.}{2000}]{York2000sdsswp}
{York} D.~G.,  et~al., 2000, \mn@doi [\aj] {10.1086/301513}, \href
  {https://ui.adsabs.harvard.edu/abs/2000AJ....120.1579Y} {120, 1579}

\bibitem[\protect\citeauthoryear{{Zee}}{{Zee}}{2003}]{Zee2003}
{Zee} A.,  2003, {Quantum field theory in a nutshell}.
Princeton University Press

\bibitem[\protect\citeauthoryear{{Zinn-Justin}}{{Zinn-Justin}}{2002}]{Zinn2002}
{Zinn-Justin} J.,  2002, { Quantum Field Theory and Critical Phenomena}.
Princeton University Press

\bibitem[\protect\citeauthoryear{{Zubeldia} \& {Challinor}}{{Zubeldia} \&
  {Challinor}}{2019}]{Zubeldia2019PlanckCMBlensing}
{Zubeldia} {\'I}.,  {Challinor} A.,  2019, \mn@doi [\mnras]
  {10.1093/mnras/stz2153}, \href
  {https://ui.adsabs.harvard.edu/abs/2019MNRAS.489..401Z} {489, 401}

\bibitem[\protect\citeauthoryear{{de Haan} et~al.,}{{de Haan}
  et~al.}{2016}]{deHann2016SPTCL}
{de Haan} T.,  et~al., 2016, \mn@doi [\apj] {10.3847/0004-637X/832/1/95}, \href
  {https://ui.adsabs.harvard.edu/abs/2016ApJ...832...95D} {832, 95}

\bibitem[\protect\citeauthoryear{{de Jong}, {Verdoes Kleijn}, {Kuijken}  \&
  {Valentijn}}{{de Jong} et~al.}{2013}]{deJong2013kidswp}
{de Jong} J. T.~A.,  {Verdoes Kleijn} G.~A.,  {Kuijken} K.~H.,   {Valentijn}
  E.~A.,  2013, \mn@doi [Experimental Astronomy] {10.1007/s10686-012-9306-1},
  \href {https://ui.adsabs.harvard.edu/abs/2013ExA....35...25D} {35, 25}

\makeatother
\end{thebibliography}
